# High temporal stability of niobium superconducting resonators by surface passivation with organophosphonate self-assembled monolayers

Harsh Gupta[*,1], Rui Pereira[2], Leon Koch[3,4], Niklas Bruckmoser[3,4], Moritz Singer[1], Benedikt Schoof[1], Manuel Kompatscher[1], Stefan Filipp[3,4], Marc Tornow[*,1,2]

[1]TUM School of Computation, Information and Technology, Department of Electrical Engineering, Technical University of Munich, 85748 Garching, Germany

[2] Fraunhofer Institute for Electronic Microsystems and Solid-State Technologies, 80686 Munich, Germany

[3]Walther-Meißner-Institut, Bayerische Akademie der Wissenschaften, 85748 Garching, Germany

[4] TUM School of Natural Sciences, Department of Physics, Technical University of Munich, 85748 Garching, Germany

*Corresponding authors- harsh.gupta@tum.de, tornow@tum.de

One main limiting factor towards achieving high coherence times in superconducting circuits is two-level system (TLS) losses. Mitigating such losses requires controlling the formation of native oxides at the metal-air interface. Here, we report the growth of alkyl-phosphonate self-assembled monolayers (SAMs) on Nb thin films following oxide removal. The impact of passivation was evaluated via the performance of coplanar waveguide resonators at 10mK, in terms of quality factor and resonant frequency, over six days of air exposure. Un-passivated resonators exhibited an ~80% increase in loss at single-photon power levels, whereas SAM-passivated resonators maintained excellent temporal stability, attributed to suppressed oxide regrowth. By employing a two-component TLS model, we discern distinct prominent loss channels for each resonator type and quantified the characteristic TLS loss of the SAMs to be ~$5\times10^{-7}$. We anticipate our passivation methodology to offer a promising route toward industrial-scale qubit fabrication, particularly where long-term device stability is critical.

# 1. Introduction

Superconducting qubits, which utilize Josephson junctions (JJs) and are capacitively coupled to coplanar waveguide (CPW) resonators, stand as promising candidates for the development of large-scale quantum computers [1–4]. The duration in which these qubits maintain their quantum mechanical state is termed as the coherence time, a critical factor affected and limited by losses, which are dominated by two-level system (TLS) losses at low temperatures and low power operation. These TLSs present in resonators, capacitors and Josephson junctions can resonantly couple to the qubits due to, e.g., atomic-scale charge oscillations, which in turn lose energy via phonon emission [4–6].

TLS losses in quantum circuits arise from defects or impurities within the materials the qubit is composed of as well as in their surrounding environment such as native oxides. Among others, they are hosted by polar impurities (like -OH) and/or oxygen bridges in native oxides formed on superconductor metal-to-air and substrate-to-air interfaces upon air exposure [7–9]. An important characteristic of TLS losses is that increasing the input power and electric field strength in the circuit potentiates the coupling to resonant TLSs, which in turn leads to a relative reduction in losses due to the saturation of TLSs at high powers. In quantum circuits, different characteristic saturation powers for TLS losses can distinguish between various loss channels. Also, this saturation enables the differentiation of power-dependent TLS losses from other, power-independent loss mechanisms at fixed temperatures.

Mitigating these losses is crucial for improving qubit coherence times. The losses can be quantified by the internal quality factor ($Q_i$) of superconducting microwave resonators. As already mentioned above, one significant loss contribution stems from the amorphous native oxide on the metal-air interfaces [10,11]. In the case of niobium, the native $NbO_x$ is commonly removed by selective wet etching, e.g., in Buffered-Oxide Etch (BOE) solution, for about 20 min. Thereby, the quality factors of resonators can be enhanced by up to one order of magnitude [10,12]. However, BOE etching alone proves inadequate in preventing the Nb surface from the subsequent regrowth of $NbO_x$. Consequently, this results in degrading quality factors over time, ultimately diminishing the performance of Nb resonators as they undergo aging. This issue is particularly crucial when aiming for sustained and uniform performance in large-scale quantum circuits.

To address this key challenge, a possible solution involves suppressing the regrowth of native oxide at the Nb-air interface by capping the freshly etched surface with a suitable passivation layer. This layer could be a thin, inorganic film with high crystalline quality contributing as little as possible with own TLSs [13]. For example, Bal et al. have recently reported the improvement of qubit coherence times by encapsulating the Nb structures with metallic thin-films from Ta or Au [14]. An alternative strategy is using organic Self-Assembled Monolayers (SAMs). SAMs comprise of a single layer of few nanometers short molecules arranged in a well-defined, ordered and closely packed, 2D-crystalline fashion [15,16]. The formation of SAMs is driven by the self-assembly properties of molecules, largely governed by van der Waals forces, hydrophobic interaction, chemisorption and hydrogen bonding [17,18]. Their tailorable properties such as wettability, electrical conductivity and chemical reactivity, make SAMs an ideal platform to prepare functional surfaces with tailored characteristics. Owing to their properties, SAMs have found applications in diverse fields [18–20], which include using SAMs as protective coatings of surfaces against oxidation or corrosion [21–23]. Well studied classes of different organic molecules, classified by their binding headgroups to the surface, include thiols, silanes, carboxylic and phosphonic acids[17,24,25].

Studies of the effect of organic coatings on the properties of conventional, element superconductors such as Nb are scarce [26,27]. Alghadeer *et al.* have recently reported the effect of passivation of Nb resonators using octatrichlorosilane molecules, with an overall increase in $Q_i$ [28,29]. However, a detailed microscopic study of the SAM characteristics as well as an investigation of the different loss channels upon passivation was not provided.

In this work, we investigate the impact of organophosphate SAM passivation on the aging and performance stabilization of Nb resonators and analyze the loss behavior of the SAMs in detail. Phosphonic acid (-$H_2PO_3$) molecules comprising, e.g., a short alkane as backbone, have been widely used for growing SAMs on different surfaces, mostly metal and semiconductor oxides, to address a wide variety of applications such as organic field effect transistors, biosensors and memristors [30–34]. We use organophosphonates here instead of the more commonly used silanes. The latter connect with only a fraction of the possible covalent bonds directly to the surface, owing to lateral cross-linking between their headgroups [35]. This may give rise to a less well-defined interface and eventually provide local sites being prone to early re-oxidation.

Our analysis focuses on the change of $Q_i$, losses and resonant frequencies of resonators over time. We systematically compare resonators with and without passivation. Additionally, our study is supported by extensive analytical techniques such as X-ray photoelectron spectroscopy (XPS), atomic force microscopy (AFM) and Fourier-transform infrared (FTIR) spectroscopy. Specifically, by using the standard TLS loss model to fit our RF data, we could identify various, distinct loss channels in both resonator types. In particular, in the course of this analysis, we could estimate the *GHz-loss of the SAM* itself, a quantity which has not been reported before, to the best of our knowledge. Such key information is fundamental to devise strategies for further reducing dielectric losses in future Nb-based quantum circuitry.

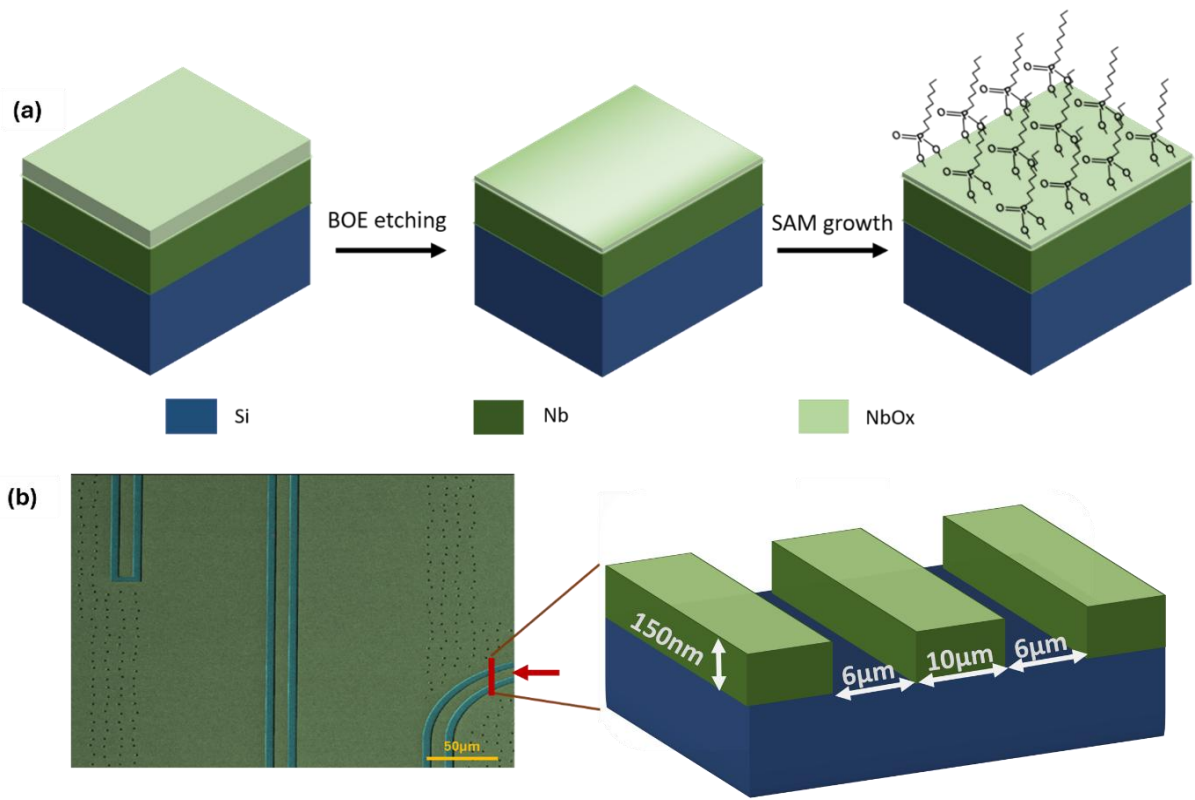

*Figure 1 (a) Schematic showing the sample processing sequence. Left: untreated sample. The Nb film is covered by a native oxide layer; Middle: 20 min BOE etched sample (termed "unpassivated"). A thin,*

*residual oxide layer is present; Right: sample after BOE etching and subsequent SAM passivation ("passivated"). (b) Left: false color SEM micrograph of part of a coplanar waveguide resonator; right: cross-sectional, 3D schematic of its geometry (not to scale). The red arrow points in the direction of the cross-sectional view.*

## 2. Results and discussion

SAM characterization:

In advance to resonator performance measurements at cryogenic temperatures, we carried out various surface characterization studies to verify the quality of our phosphonate monolayer coatings on Nb. As organophosphonate precursor molecules we choose decylphosphonic acids. To estimate the thickness of our SAMs grown on planar Nb films post 20 min BOE etching, ellipsometry measurements were conducted which yielded a measured film thickness of 1.2±0.2 nm (Supplementary figure S3). This agrees well with the expected thickness of a molecular monolayer of slightly tilted C10 alkyl chains, whose contour length are about 1.3 nm. Water contact angle measurements were carried out to compare the Nb surface wettability before and after coating with a SAM. Right after BOE etching, Nb stripped of (most of) its native oxide, exhibits a contact angle of about 42°±2° (Figure 2a). Such a contact angle indicates a rather hydrophilic surface, which can be expected for a thin, clean residual oxide layer, that is most likely still present on the Nb film after BOE etching: either, the oxide was not completely etched or already slightly regrown upon exposure to air post BOE etching. In contrast, Nb passivated with a phosphonic acid SAM post BOE etching reveals a significantly higher contact angle of 103°±1° (Figure 2b). This rather hydrophobic surface suggests the successful formation of a well-ordered aliphatic SAM, whose surface termination by methyl groups typically leads to contact angles in excess of 100° [36–40]. Contact angle stability (figure2c) indicates suppression of oxide regrowth. The high contact angle of SAM-passivated Nb remains unchanged over 14 days, reflecting durable hydrophobicity and resistance to air-induced oxidation. Unpassivated Nb shows minimal variation, indicating that the nature of the thin oxide on the BOE-etched Nb thin film does practically not change upon aging.

AFM measurements were taken before and after SAM growth, with root mean squared (RMS) roughness values for Nb and SAM passivated Nb thin films of 0.5± 0.1nm and 0.6± 0.1nm, respectively, see, Figure 2e, f). This indicates the uniform and conformal growth of phosphonic acid SAMs on Nb. FTIR measurements for the SAM-coated sample showed absorption peaks in the wavenumber range 2800 – 3000 cm$^{-1}$, characteristic for C-H stretching vibrations of hydrocarbons (Figure 2d). The position of the $CH_{2,s}$ and $CH_{2,a}$ peaks in case of our decylphosphonic acid SAM under study are 2851 cm$^{-1}$ and 2919 cm$^{-1}$, respectively, hence indicating a highly ordered SAM on the Nb surface [24,36,40].

Resonator measurements:

Microwave measurements were carried out at ~10 mK on unpassivated and passivated niobium CPW resonators. The internal quality factors $Q_i$ of the resonators were extracted from $Q_l$ (loaded quality factor) and $Q_c$ (coupling quality factor) (for detailed analysis, see supplementary information: resonator measurement) using the following relation [10,41]:

$$\delta_i = \frac{1}{Q_i} = \frac{1}{Q_l} - \frac{1}{Q_c} \qquad (1)$$

By design, our resonators operate in close to the critical coupling regime, i.e., $Q_c \sim Q_l$. The resonators´ performance is characterized by the internal quality factor $Q_i$, which is related to their internal losses $\delta_i$. To investigate the effect of the resonators´ aging on their $Q_i$ values, measurements were performed with both, the passivated (BOE etched and coated with SAM) and unpassivated (BOE etched, only) resonators on day 0 (freshly prepared resonators), day 2 (aging in air for ~48 hours), and day 6 (aging in air for ~144 hours).

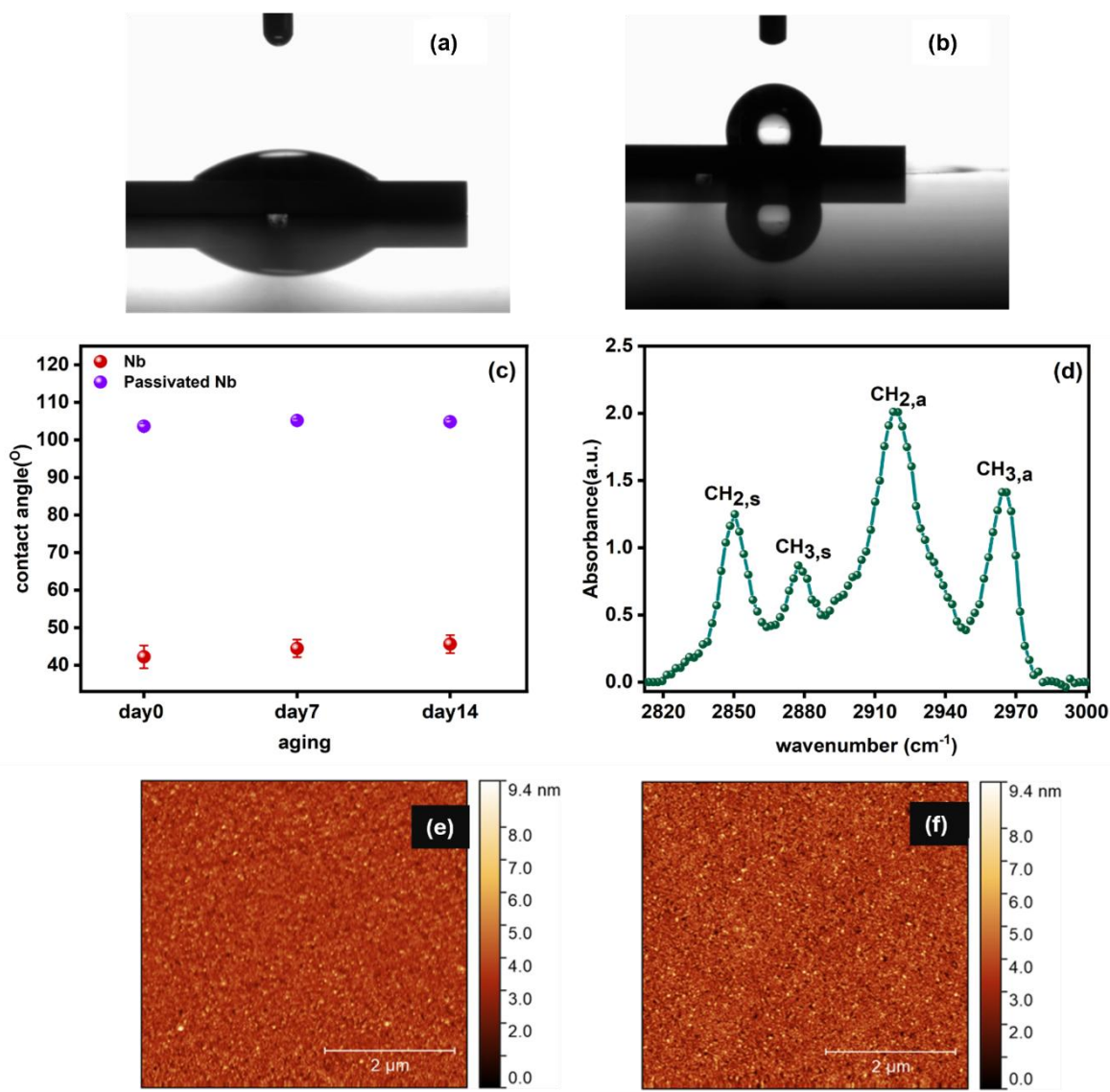

*Figure 2: Sample surface characterization. (a) Camera photo during contact angle (CA) measurement on freshly BOE-etched, planar Nb, (b) CA measurement on BOE etched, decylphosphonic acid SAM-passivated, planar Nb, not aged (day0). A significant increase in contact angle (increased hydrophobicity) post SAM growth is evident. (c) Temporal development of the contact angle for planar Nb samples as in a, b). The high contact angle values (purple dots) of the passivated sample remain practically unchanged with aging over 14 days. Error bars show the standard deviation of the measurement (note that, the error of the passivated Nb-data is too small to be visible). (d) FTIR spectra for passivated Nb showing the presence of different vibrational modes of alkyl molecules on the surface. Labels at the peaks indicate: the symmetric stretch vibration ($CH_{2,s}$) and the asymmetric stretch vibration ($CH_{2,a}$) of the methylene-groups, as well as the symmetric stretch mode ($CH_{3,S}$) and the asymmetric stretch mode ($CH_{3,a}$) of the terminating methyl groups of the molecules. (e), (f) AFM images*

*showing the topography of an unpassivated Nb film after BOE treatment with RMS roughness 0.5±0.1 nm and a decylphosphonate-passivated Nb thin film with RMS roughness 0.6±0.1 nm, respectively.*

Firstly, we will assess the development of the $Q_i$ values of the un-passivated resonators upon aging. $Q_i$ values averaged over all resonators (with different frequencies) on the same chip are presented in Figure 3. For all these resonators, the measured $Q_i$ values increased with increasing incident power, in the range from $10^{-1}$ to $10^8$ average photon numbers, which is characteristic of TLS losses in the resonators. These mean values decrease upon aging, signifying an increase in overall internal losses. The $Q_i$ vs. $<n>$ trace shows no saturation for high powers, a behavior which is uncharacteristic for a single contributing TLS type, hence indicating the presence of multiple TLS loss channels. Consequently, to analyze in detail the $Q_i$ versus photon number $<n>$ (power) dependence, the power dependence of $Q_i$ measured for each resonator was fitted with a multi-TLS loss expression [12,42–45]:

$$\tan(\delta_i) \approx \delta_i = \frac{1}{Q_i} = \sum_i \frac{\tilde{\delta}_{i,TLS}}{\left[1+\left(\frac{n}{n_i}\right)\right]^{b_i}} + \delta^0 \qquad (2)$$

where $\tilde{\delta}_{i,TLS}$ is the weighted characteristic TLS loss for loss channel $i$, according to $\tilde{\delta}_{i,TLS}=p_i \cdot \delta_{i,TLS}(T)$, with $p_i$ the geometry and dielectric constant-dependent participation ratio of the $i^{th}$ volume and $\delta_{i,TLS}(T)$ the corresponding, material-specific loss tangent. $\delta^0$ is the overall (non-TLS) power-independent loss, $n$ is the photon number according to the input power, $n_i$ is the critical photon number related to the saturation field of the corresponding TLSs, and $b_i$ is a phenomenological parameter quantifying the deviation from the standard TLS loss model, with typical values $0 < b_i < 0.5$. Here, lower values of $b_i$ indicate a stronger mutual interaction of TLSs [12,46–48].

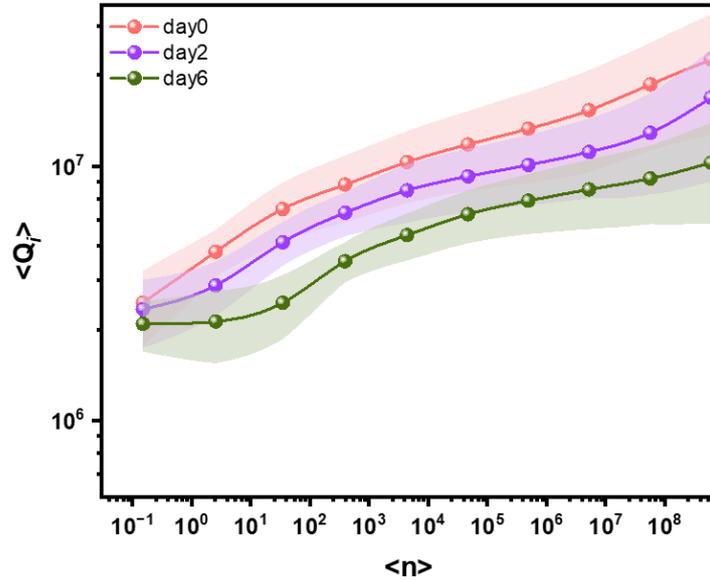

**Figure 3**: *Mean internal quality factor $Q_i$ for un-passivated Nb resonators against mean photon number, for different stages of aging: day 0 (freshly etched), day2 and day6. The mean values were obtained by averaging over the $Q_i$ values of resonators with resonance frequencies ranging from ~4-8GHz. The shaded region in the plot denotes the standard deviation of the data. With aging in air, $Q_i$ decreases significantly for un-passivated Nb resonators throughout the entire photon number range under investigation. Measurements taken at T~10 mK.*

To describe the behavior of the $Q_i$ values vs photon number, two different (distinct) TLS loss channels, i.e., $i=1,2$ in equation 2, were considered. Since no saturation of $Q_i$ was observed in the high photon number limit (see the plots of Figure 3), the term corresponding to the power independent losses $\delta^0$ was neglected in the fitting procedure. However, this does not affect the values obtained from the fittings for the other parameters of equation 2. The data was fitted within a 95% confidence interval, with constraints of $b_i < 0.6$, $n_1 < 10$ and $n_2 > 10^3$ for the $Q_i$-$n$ plot, with no strong correlation (<0.7) between different parameters (for typical behavior, see Supplementary figures S9, S10). All the fitting parameters, namely the phenomenological parameter $b_i$, the critical photon number $n_i$ and characteristic losses $\tilde{\delta}_{i,TLS}$ were then derived from the fitting routine. Details of the distribution of $n_i$ and $b_i$ over the different resonator frequencies on the chip (for day0), of the change of their mean values upon aging, and of the change of $\tilde{\delta}_{1,TLS}$ and $\tilde{\delta}_{2,TLS}$ upon aging, are all displayed in Figure 4.

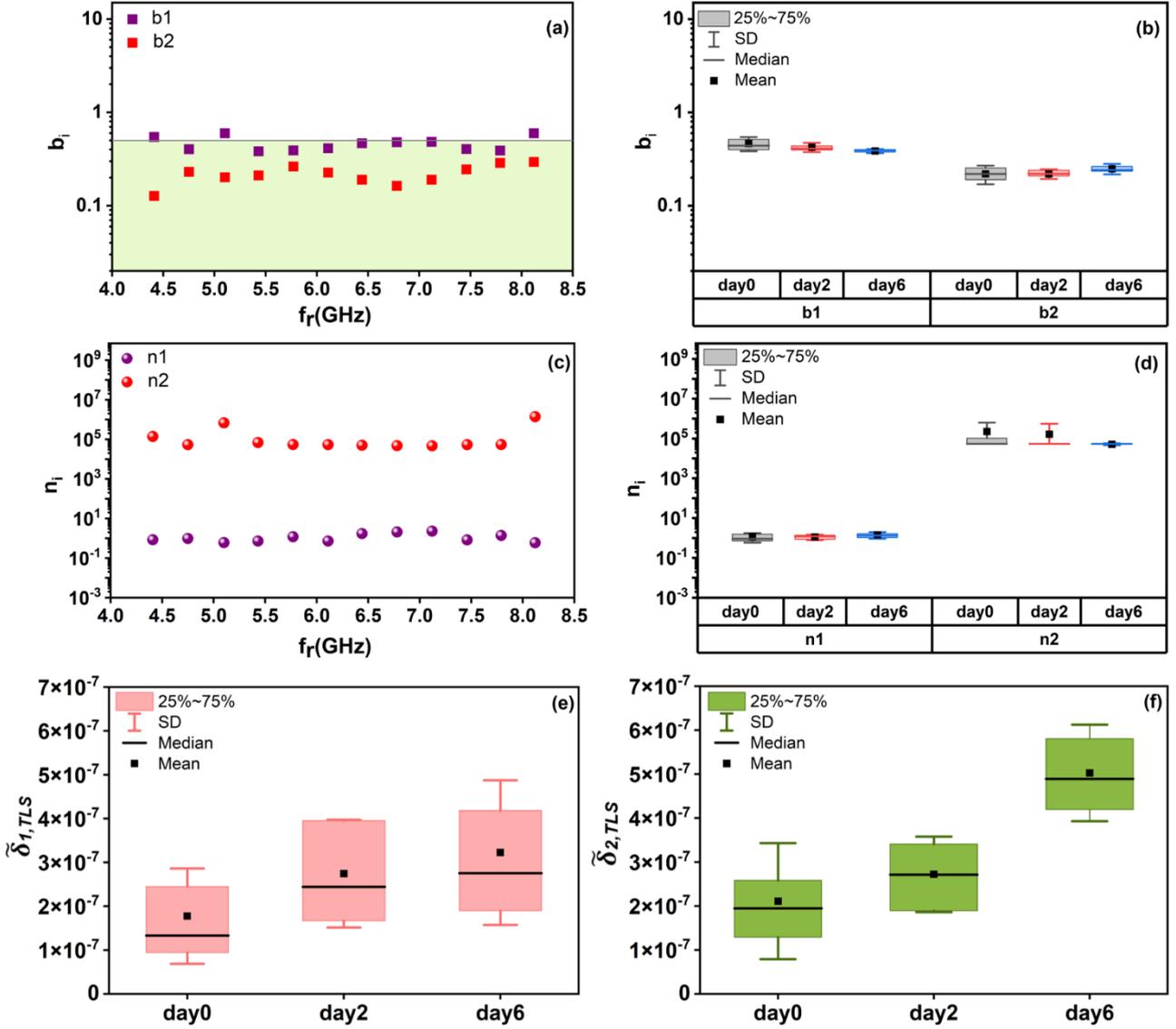

*Figure 4: TLS model fitting parameters, according to eq. 2, for un-passivated resonators: (a) parameters $b_1$, $b_2$ for day0, as function of resonator frequency. The green shaded area denotes the region below 0.5, (b) Variation of $b_1$, $b_2$ with aging over 6 days; in all box plots the color-filled box ranges from the first to the third quartile, the solid point and line represent the mean and the median, respectively, whiskers denote the standard deviation (SD). Both, $b_1$ and $b_2$ stay mostly below 0.5, signifying dominating interacting TLSs. (c) Critical photon number $n_1$ (order of ~$10^0$), $n_2$ (~$10^5$) for two distinct TLS channels for day0, as function of resonator frequency, (d) variation of $n_1$, $n_2$ with aging over 6 days. The two characteristic photon numbers do not vary significantly upon aging. (e), (f)*

*Variation of characteristic losses $\tilde{\delta}_{1,TLS}$ and $\tilde{\delta}_{2,TLS}$, respectively, for the two distinct TLS loss channels as function of aging time. A clear increase indicates the accumulation of losses corresponding to both TLS channels upon aging in air over 6 days.*

For quick reference and direct comparison, the mean values of $n_i$, $b_i$ and $\tilde{\delta}_{i,TLS}$ are also summarized in Table 1, for both, day0 and day6. As apparent from this composition, the values obtained for $b_1$ and $b_2$ are typically lower than 0.5, with $b_2$ being notably lower for all different resonators on the investigated chip (Figure 4a). This suggests interacting TLSs for both loss channels, with the channel ($i=2$) ($b_2$) featuring even more interacting TLSs in nature. By employing the two component-TLS model we identify distinct significant loss channels as denoted by strongly dissimilar critical photon numbers in the range $n_1 \sim 10^0$ and $n_2 \sim 10^5$, cf. Figure 4c.

*Table 1: Mean fitting parameters extracted from fitting the two-TLS model to the data of the un-passivated Nb resonators, for day0 (fresh) and day6 (aged).*

| Parameter | day0 | day6 |
| --- | --- | --- |
| $\langle b_1 \rangle$ | 0.46 | 0.38 |
| $\langle b_2 \rangle$ | 0.21 | 0.24 |
| $\langle n_1 \rangle$ | 1.16 | 1.45 |
| $\langle n_2 \rangle$ | $2.2 \times 10^5$ | $5.1 \times 10^4$ |
| $\langle \tilde{\delta}_{1,TLS} \rangle$ | $1.7 \times 10^{-7}$ | $3.2 \times 10^{-7}$ |
| $\langle \tilde{\delta}_{2,TLS} \rangle$ | $2.1 \times 10^{-7}$ | $5.0 \times 10^{-7}$ |

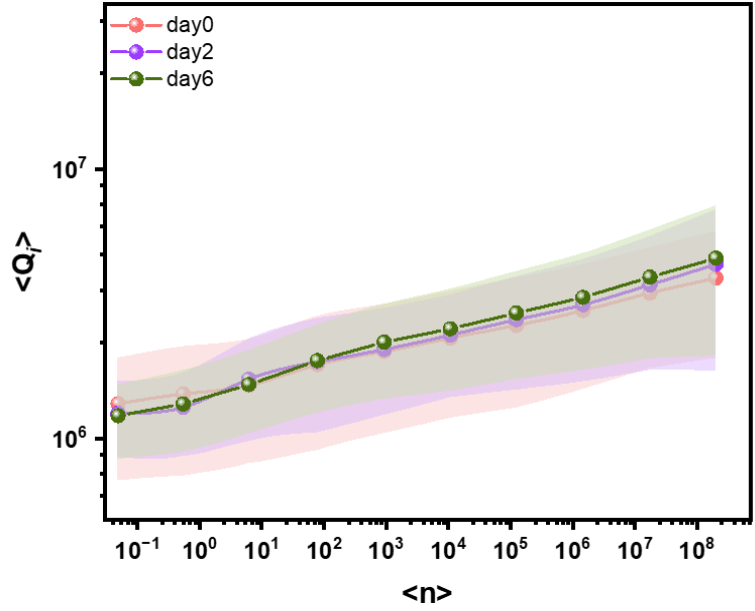

*Figure 5*: Mean internal quality factor ($Q_i$) for SAM-passivated Nb resonators against mean photon number, for different stages of aging: day 0 (freshly etched), day2 and day6. The shaded regions in the plot denote the standard deviation of the data. With aging in air, $Q_i$ remains practically unchanged for all resonators throughout the entire photon number range under investigation.

TLS losses corresponding to $n_1$ can be ascribed to overall intrinsic losses of the resonators and will be discussed further below. In contrast, as we will argue in the context of the data for the passivated samples below, we assign the loss channel corresponding to $n_2$ to losses within $NbO_x$, primarily $Nb_2O_5$, which starts to regrow upon exposure to air (aging) [12]. The degradation in $Q_i$ values observed upon aging for un-passivated resonators can be understood in terms of the contributing, characteristic TLS losses ($\tilde{\delta}_{i,TLS}$). Figure 4e, f) show the variation of these losses, corresponding to two different loss channels in unpassivated resonators, with aging. Both characteristic losses $\tilde{\delta}_{1,TLS}$ and $\tilde{\delta}_{2,TLS}$ increase, with $\tilde{\delta}_{2,TLS}$ even more significantly, thus leading to the degradation in the $Q_i$ values, upon aging for 6 days. This notable surge in $\tilde{\delta}_{2,TLS}$ losses coincides with the accumulation of TLSs with the growth of $NbO_x$ on the BOE etched Nb surface. We observe that the $\tilde{\delta}_{1,TLS}$ and $\tilde{\delta}_{2,TLS}$ variation with aging is significantly larger than the on-chip resonator-to-resonator variation

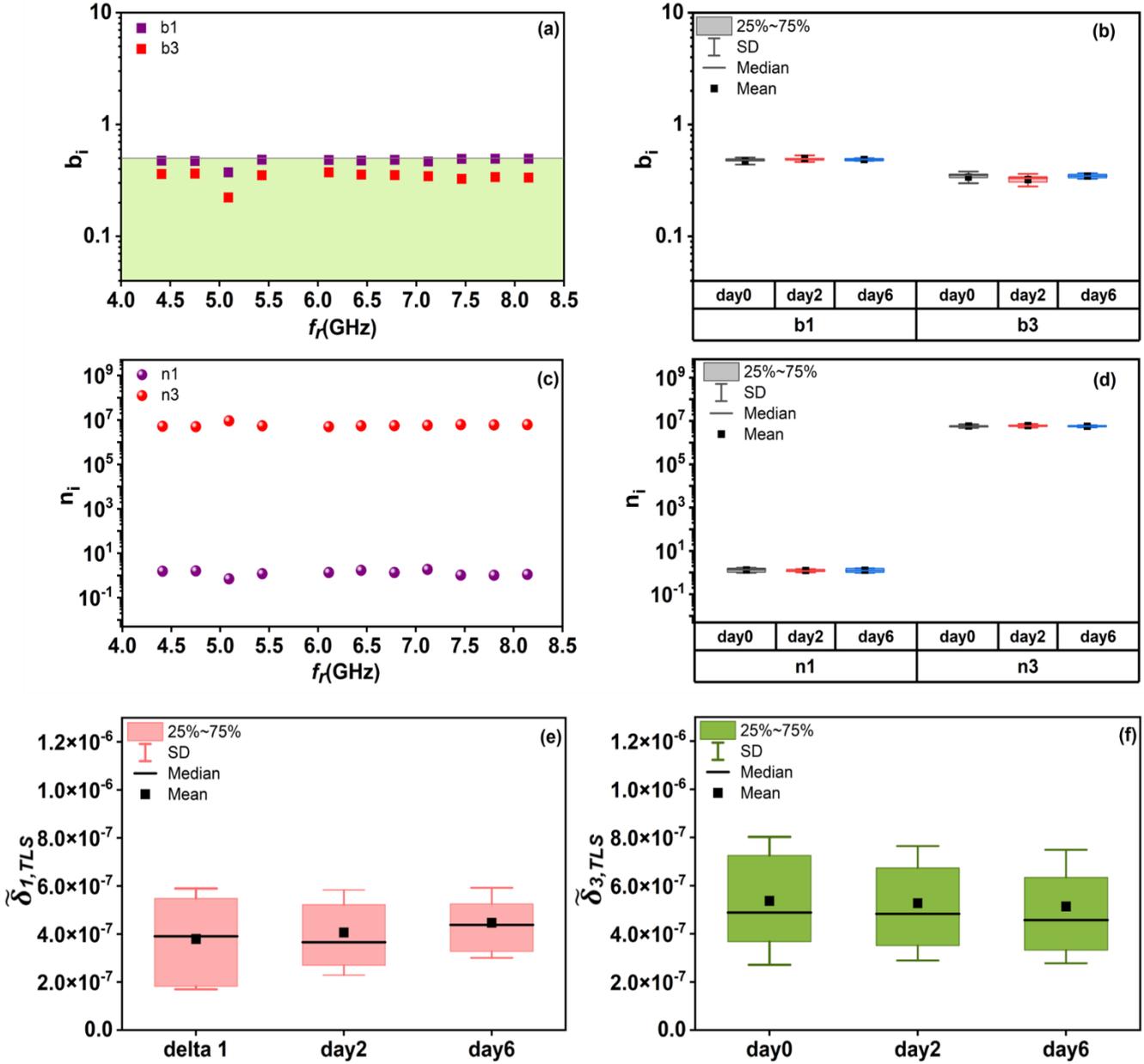

*Figure 6*: TLS fitting parameters for SAM passivated resonators (a) parameters $b_1$, $b_3$ for day0, (b) variation of $b_1$, $b_3$ with aging over 6 days. Both $b_1$, $b_3$ are close to or below 0.5 signifying mostly interacting TLSs. (c) Critical photon numbers $n_1$ (~$10^0$), $n_3$ (~$10^7$) for two distinct TLS channels for the resonators at day0. (d) variation of $n_1$, $n_3$ with aging over 6 days. The two characteristic photon numbers do not vary upon aging. (e), (f) Variation of characteristic losses $\tilde{\delta}_{1,TLS}$ and $\tilde{\delta}_{3,TLS}$, respectively, for two distinct TLS loss channels upon aging. Compared to the un-passivated resonators (Fig. 4), a clear stabilization of losses corresponding to both TLS channels against aging in air over 6 days is evident.

The $Q_i$ values of the passivated (20 min BOE etched and SAM grown) Nb resonators were measured in the same way at ~10 mK for the three aging stages (day 0, day 2 and day 6); their dependence on photon number is shown in Figure 5. As can be seen, no degradation of the resonators with aging is observed for the passivated resonators. Analogous to the un-passivated resonators, the dependence of $Q_i$ versus $<n>$ measured for each different resonator was fitted with equation 2, and the fitting results for $n_i$, $b_i$ and $\tilde{\delta}_{i,TLS}$ are displayed in Figure 6, together with a summary of selected mean values in Table 2. In fact, the data was fitted with the same constraints as for un-passivated resonators within a 95% confidence interval, and again for two channels. While fitting the TLS model to the data confirms once more the presence of two discrete loss channels in the resonators, we observe a distinct change with respect to the un-passivated case: while about the same critical photon numbers $n_1 \sim 10^0$ were obtained as for the un-passivated resonators (relating them to the same corresponding, intrinsic TLS losses in the single photon region), the critical photon numbers for the second loss channel were significantly higher, $\sim 10^7$.

*Table 2: Mean fitting parameters extracted from fitting the two-TLS model to the data of the passivated Nb resonators, for day0 (fresh) and day6 (aged).*

| Parameter | **Day0** | **Day6** |
|---|---|---|
| $<b_1>$ | 0.47 | 0.48 |
| $<b_3>$ | 0.33 | 0.34 |
| $<n_1>$ | 1.32 | 1.27 |
| $<n_3>$ | $6.8 \times 10^6$ | $7.0 \times 10^6$ |
| $<\tilde{\delta}_{1,TLS}>$ | $3.8 \times 10^{-7}$ | $5.3 \times 10^{-7}$ |
| $<\tilde{\delta}_{3,TLS}>$ | $4.4 \times 10^{-7}$ | $5.2 \times 10^{-7}$ |

For this reason, we labelled its parameters with index "3" (see, Figure 6 and Table 2) and conclude that it must be different from loss channel no. 2. At the same time, the previously identified loss channel no. 2 can now – in retrospective – be ascribed to NbOx, as this layer has mostly been removed for the passivated samples and is therefore expected to contribute significantly less to the overall distribution of loss channels here. It should be noted that contributions from NbOx likely still exist. However, they are anticipated to be significantly lower than the loss contributions assigned to channel no. 3, thereby largely masked by the latter in the passivated resonators.

Upon aging, the characteristic, mean internal losses $\tilde{\delta}_{1,TLS}$ increase slightly, however still within the variance of the day0 data, where $\tilde{\delta}_{3,TLS}$ shows no significant change (Figure 6e, f). The values obtained for the parameters $b_1$ and $b_3$ are typically close to or lower than 0.5, with $b_3$ being marginally lower than $b_1$ across all resonators (Figure 6a), suggesting a predominantly interacting nature of TLSs for both loss channels – similar as for the un-passivated case.

Examining the TLS model for passivated and unpassivated resonators highlights distinct interfacial loss mechanisms and performance evolution under ambient conditions. Both resonator types share a stable dominant channel in the single photon regime ($n \sim 10^0$), represented by the value $n_1$, unchanged over time (Fig. 7), suggesting intrinsic losses from Nb growth, patterning, and substrate-related interfaces (metal–substrate, air–substrate) are the primary contributors. We therefore globally assign these intrinsic losses as $\tilde{\delta}_{1,TLS} = \tilde{\delta}_{int.}$. It should be noted that the $\tilde{\delta}_{int.}$ for passivated resonators are somewhat higher than for un-passivated ones. This can be due to losses additionally incorporated during the SAM growth process, possibly due to the adverse effect of aging in solvent for 48hrs and left-over solvent traces after SAM deposition. For un-passivated resonators, the losses $\tilde{\delta}_{2,TLS}$ corresponding to the second loss channel, characterized by a higher critical photon number ($n_2 \sim 10^5$), exhibit a significant increase with aging. Since the thickness of the oxide at the Nb surface (primarily $Nb_2O_5$) is the main parameter that changes as a function of aging, we conclude – as already noted earlier – that this loss component is linked to regrown $Nb_2O_5$ ($\tilde{\delta}_{2,TLS}= \tilde{\delta}_{Nb2O5}$).

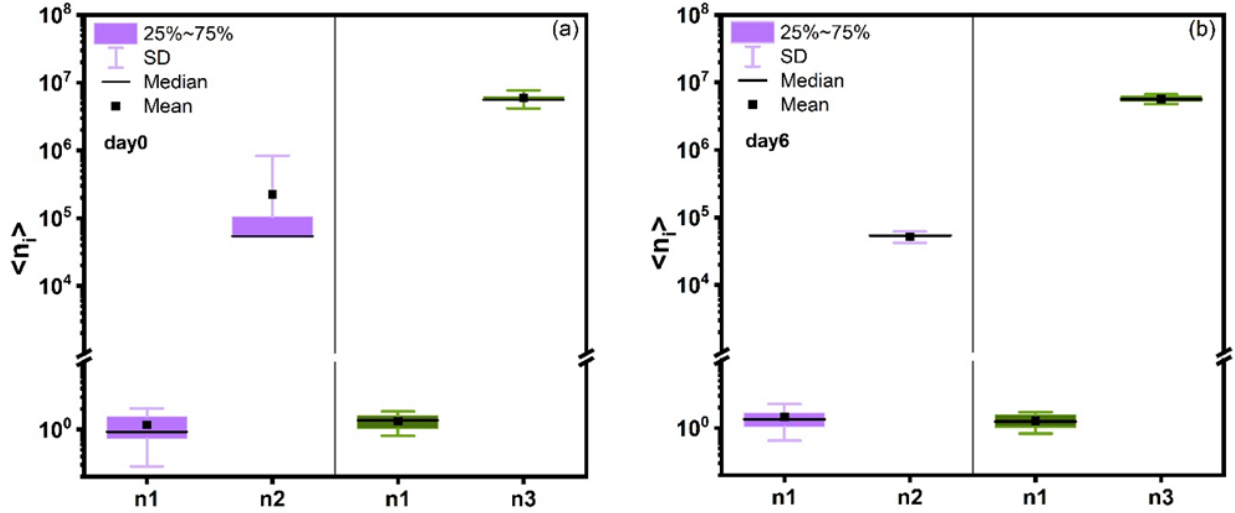

*Figure 7: Characteristic critical photon numbers extracted from the TLS model: $n_1$, $n_2$ for un-passivated resonators (purple) and $n_1$, $n_3$ for passivated resonators (green) (a) variation over different resonant frequencies on day 0 (b) variation on day 6. In the single photon region, a common loss channel ($n_1$) is observed whereas, two distinct loss channels for un-passivated and passivated resonators, corresponding to $n_2$ and $n_3$, respectively, are noted (both in high photon number regime).*

Conversely, in passivated resonators, the second loss channel's $n_3 \sim 10^7$ notably differs from $n_2$ observed in un-passivated resonators (Figure 7). Importantly, the $\tilde{\delta}_{3,TLS}$ associated with this high critical photon number does not change during aging, reflected in the high temporal stability of SAM-passivated resonators (Figure 6f). The SAM is the additional layer in this case; it is expected to remain stable for even longer times than the total aging period (6 days), as also suggested by contact angle measurements carried out up to 14 days (Figure 2c). This loss channel corresponding to a high critical photon number ($n_3 \sim 10^7$) is therefore ascribed to the passivating SAM itself ($\tilde{\delta}_{3,TLS} = \tilde{\delta}_{SAM}$). The loss $\tilde{\delta}_{SAM}$, corresponding to the decylphosphonic acid SAM on Nb, as extracted from the loss model, was estimated to be $\approx 5 \times 10^{-7}$ (see, Figure6f). Taking into account typical ranges for thin-film participation ratios at the metal-air interface, $10^{-3} - 10^{-4}$ [49], the corresponding material value $\tilde{\delta}_{SAM}$ is in the similar range as reported for common inorganic thin-film dielectrics [50–52] and for the analogue (bulk) organic polymer polyethylene [53]. To the best of our knowledge, this is the first time that the dielectric loss of a SAM in the GHz regime has been reported.

The impact of the SAM passivation on the temporal performance of Nb resonators is noteworthy. In figure 8, we present a comparative analysis of the resonators' performance over the 6-days aging period, focusing on the relative variation of losses. Un-passivated resonators exhibit a substantial rise in losses, with $\tilde{\delta}_{int.}$ increasing by 80 %, which may be attributed to losses on the open Si surface with re-growing oxide, and $\tilde{\delta}_{Nb2O5}$ by even 135%, after 6 days – certainly due to the regrowth of oxide. In contrast, passivated resonators show no significant increase in losses compared to their counterpart during aging: $\tilde{\delta}_{int.}$ increases by only ~16%, possibly due to the open Si surface, which is more prone to re-oxidation as phosphonate SAMs would show a poor growth on H-terminated silicon. $\tilde{\delta}_{SAM}$ even show a minor reduction, which we regard as not significant within the uncertainty of the measurement.

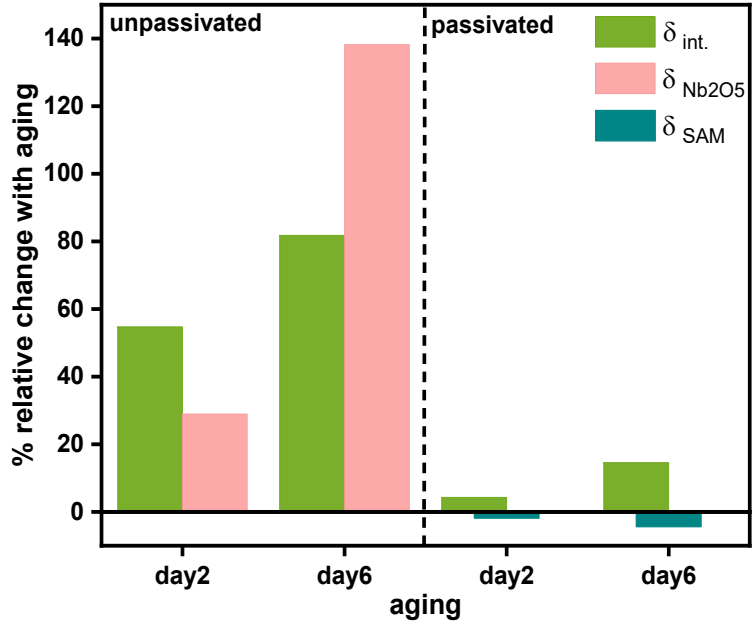

*Figure 8*: *Relative changes in different losses upon aging in air. A significant increase in losses, both $\delta_{int}$ and $\delta_{Nb2O5}$, for un-passivated resonators (left) is evident, whereas passivated resonators (right) show excellent temporal stability with aging.*

In summary, these findings are well in agreement with the expectation that native oxides constitute a primary source of losses in resonators [5,10,12,46]. When these oxides grow in an unhindered manner, the related losses grow significantly as well. In turn, when the oxide growth is prevented by a suitable passivation layer, and this passivation layer (the SAM) is not subject to any own degradation with time, losses are well stabilized.

Besides achieving and maintaining high resonator quality factors, also the stability of their resonance frequencies is another critical factor in superconducting qubit circuitry. Here in particular, as the coupling of TLSs with the resonators introduces supplementary energy dissipation and damping effects, relative shifts in resonance frequencies typically occur concomitantly with the degradation of the quality factors (Supplementary figure S8). Figure 9 displays the measured change in resonance frequencies for un-passivated and passivated resonators, after 2 and 6 days of aging. In fact, both resonators exhibit a characteristic shift of resonance frequencies upon aging, in the order of a few 100 kHz, while this change is notably larger for the un-passivated resonators (Figure 9a).

Specifically, after 6 days of aging in ambient air, we observed that un-passivated resonators showed a 225% higher mean resonator frequency shift compared to the passivated ones. This can be attributed to the increased number of TLS, resulting in a change of the dielectric function at the Nb surface hosting the TLSs. The relative shift in resonant frequency due to the coupling of TLSs with the resonators is given by equation 3 [5,10,54,55]

$$\frac{\Delta f_r}{f_r} = \frac{\widetilde{\delta}_{TLS}}{\pi}\left[Re\ \Psi\left(\frac{1}{2} + \frac{hf_r}{ik_BT}\right) - \log\left(\frac{hf_r}{k_BT}\right)\right] \sim \frac{\Delta\epsilon}{\epsilon} \qquad (3)$$

where $\Psi$ is the complex digamma function, $f_r$ is the resonant frequency, $\epsilon$ is the dielectric function and $\tilde{\delta}_{TLS}$ is the weighed characteristic dielectric TLS loss. At constant temperature, the magnitude of relative change in frequencies is proportional to the TLS density as, $\frac{\Delta f_r}{f_r} \propto \tilde{\delta}_{TLS}$. In our case, the shift in frequencies is related to the change in total TLS losses ($\Delta\tilde{\delta}_{TLS}=\Sigma(\Delta\tilde{\delta}_i)$, i.e., the increase), originating from all different loss channels. The substantial frequency shift observed in un-passivated resonators is linked to the notable increase in total TLS losses associated with the loss channels $\tilde{\delta}_{int.}$ and $\tilde{\delta}_{Nb2O5}$, upon aging (Figure 9b).

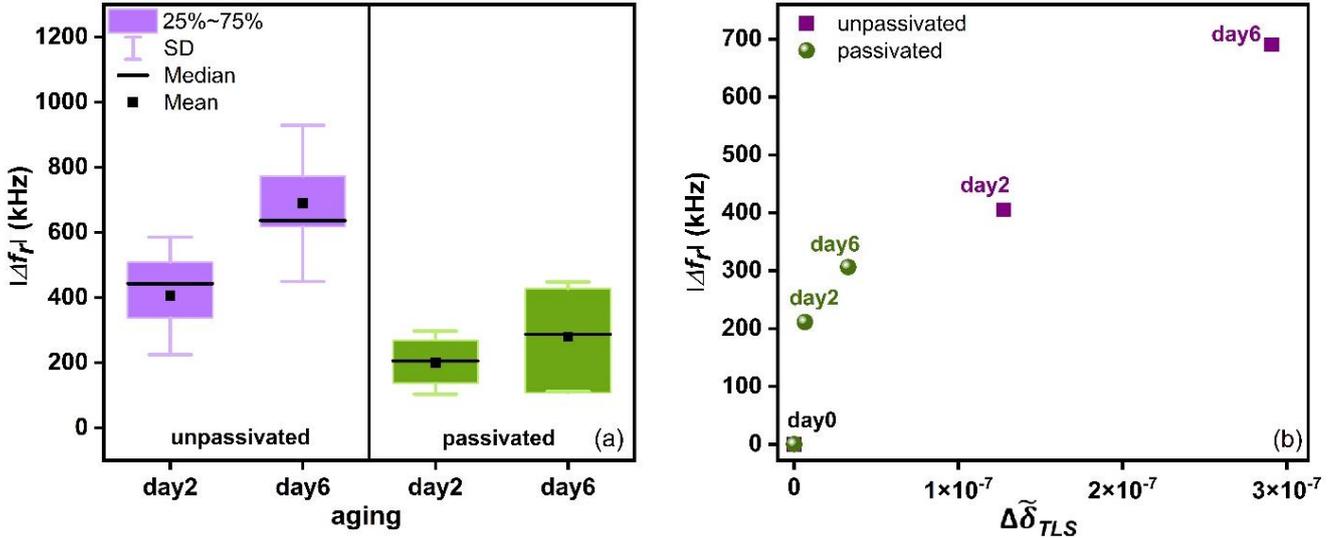

*Figure 9:* (a) Statistical *variation of resonant frequencies (absolute values) with aging for: un-passivated (purple) and passivated (green) resonators. A significantly higher shift in resonant frequencies for un-passivated resonators is evident. (b) Mean shift in resonant frequencies for un-passivated and passivated resonators against the total change in their TLS losses with aging.*

In contrast, the resonance frequencies of passivated resonators show a better stability, with relative changes in total losses mainly assigned to $\tilde{\delta}_{int.}$, as $\tilde{\delta}_{SAM}$ was found to almost not change over time. The mean relative frequency shift for passivated resonators (approx. 300kHz) even after 6 days of aging is still less than the one for un-passivated resonators (approx. 400kHz) after only 2 days of aging.

We assign the prolonged temporal stability of both, the quality factor and the resonance frequency, for SAM-passivated resonators to their surfaces' resilience against the re-growth of native oxide during exposure to air. This hindrance to native oxide formation can be attributed to the enhanced and enduring sealing of the surface against chemical reactions, which has been used as one of the initial applications of SAMs: corrosion protection [22,56]. The bare sealing is strongly supported by the distinct hydrophobicity of the $CH_3$-terminated SAM surface (see, Figure 2), preventing the formation of various loss-inducing polar impurities including –OH species, which would otherwise accumulate on the Nb surface and contribute to oxide regrowth as well as to increased losses on their own [8,11].

X-ray photoelectron spectroscopy:

Finally, to independently verify our hypothesis of hindered oxide regrowth, we investigated the surface chemical composition by carrying out a detailed XPS study, the results of which are summarized in Figure 10. Specifically, we probed both, un-passivated and passivated planar Nb thin-film samples, aged for 6 days, and the constituent elements were then identified from survey spectra. The presence of organophosphate molecules on the passivated Nb films was confirmed with high-resolution P-2p spectra (Supplementary figure S4). For high-resolution Nb 3d spectra, post Shirley background correction and fitting, we observe the three apparent, pairs of symmetric peaks to the various oxidation states of Nb and one asymmetric pair of peaks to the niobium elemental metal (Nb$^0$) (3d$_{5/2}$ component at 202.01eV). We can attribute the Nb$^{2+}$, Nb$^{4+}$ and Nb$^{5+}$ oxidation states to the symmetric pair of peaks, corresponding to NbO, NbO$_2$ and Nb$_2$O$_5$, to the 3d$_{5/2}$ component positioned at 203.3eV, 205.8eV and 207.3eV, respectively (Figures 10a, b) [12,57–59]. Each state is split into two peaks due to the strong spin-orbit coupling in niobium. One notable observation is the pronounced damping of the Nb$^0$ signal for the un-passivated Nb, when compared to passivated Nb, confirming the growth of a thicker oxide on the un-passivated sample. This can be further understood in terms of the relative contribution from Nb$^0$ in the Nb 3d spectra (Figure 10c): a higher relative contribution (~50%) is observed for passivated Nb, whereas the contribution in un-passivated Nb is significantly lower (~16%), due to damping of the Nb signal by the thicker oxide. It should be noted that the overall Nb intensity is also reduced for the passivated films due to damping by the SAM; hence, we here consider only the relative contributions of the different components from the Nb 3d spectra.

The thicknesses of the different oxide components were extracted from the XPS spectra using a straight line-effective attenuation length model, suggesting Nb$_2$O$_5$ as the most significant oxide on Nb [10,58,60] (Supplementary Information: X-ray photoelectron spectroscopy). The result of this estimation is shown in Figure 10 d). Here, the total oxide thickness for un-passivated Nb is estimated to be ~3.8 nm, whereas the passivated Nb shows an overall oxide thickness of ~1.2 nm. These values are in close agreement with the thickness values obtained by ellipsometry (Supplementary table S1). It should be noted that the Nb oxide thickness on passivated Nb may be overestimated, as a significant contribution for the Nb$^{+2}$ (NbO) and Nb$^{+4}$ (NbO$_2$) is expected to arise from the bonding of phosphonic acid molecules via their -H$_2$PO$_3$ head groups on Nb. The contribution from Nb$^{+5}$ in passivated Nb is expected to arise from leftover oxide after BOE etching. In addition to the analysis of the Nb 3d spectra, we also recorded and evaluated the valence band (VB) spectra using XPS near the Fermi level ($E_f$= 0 eV). Figures 10 (e, f) represent the VB spectra including both the Nb 4d–O 2p valence bands for un-passivated and passivated Nb films post 6-day aging after background subtraction as well as the P 3p states for passivated thin films. For metallic niobium, the 4d states spread up to 4eV with a maximum close to 1eV. In case of the p–d band states, different O 2p-Nb 4d hybridized states corresponding to NbO, NbO$_2$ and Nb$_2$O$_5$ result in a wide peak in the energy region of 3.5 eV to 9 eV below $E_f$ [61,62] (Supplementary figure S5). All the oxide Nb 4d-O 2p states completely disappear in the vicinity of $E_f$. In case of passivated films we could observe the phosphorus 3p signature around 3eV (corresponds to either P-O or P:H), shape and width of which can be affected by hybridization with O2p states [63,64]. VB spectra brings more details regarding the effect of SAM passivation on the further oxidation of Nb films: no significant change in states is observed in the vicinity of $E_f$, with a small shift of the Nb 4d peak for the passivated film, only.

However, the metallic Nb signal (Nb 4d states) is relatively stronger in passivated films in comparison to un-passivated thin films. On the other hand, the oxide contribution in un-passivated films is significantly higher than in passivated films, indicating a lower density of states for the Nb 4d-O 2p states corresponding to oxides for the passivated film, i.e., a thinner oxide on passivated films as indicated by the XPS core spectra (Figure 10 a,b) as well. This further affirms the ability of SAMs to restrict the regrowth of oxide on the Nb surface, which in turn agrees well with the observed stability over time, for the passivated resonators.

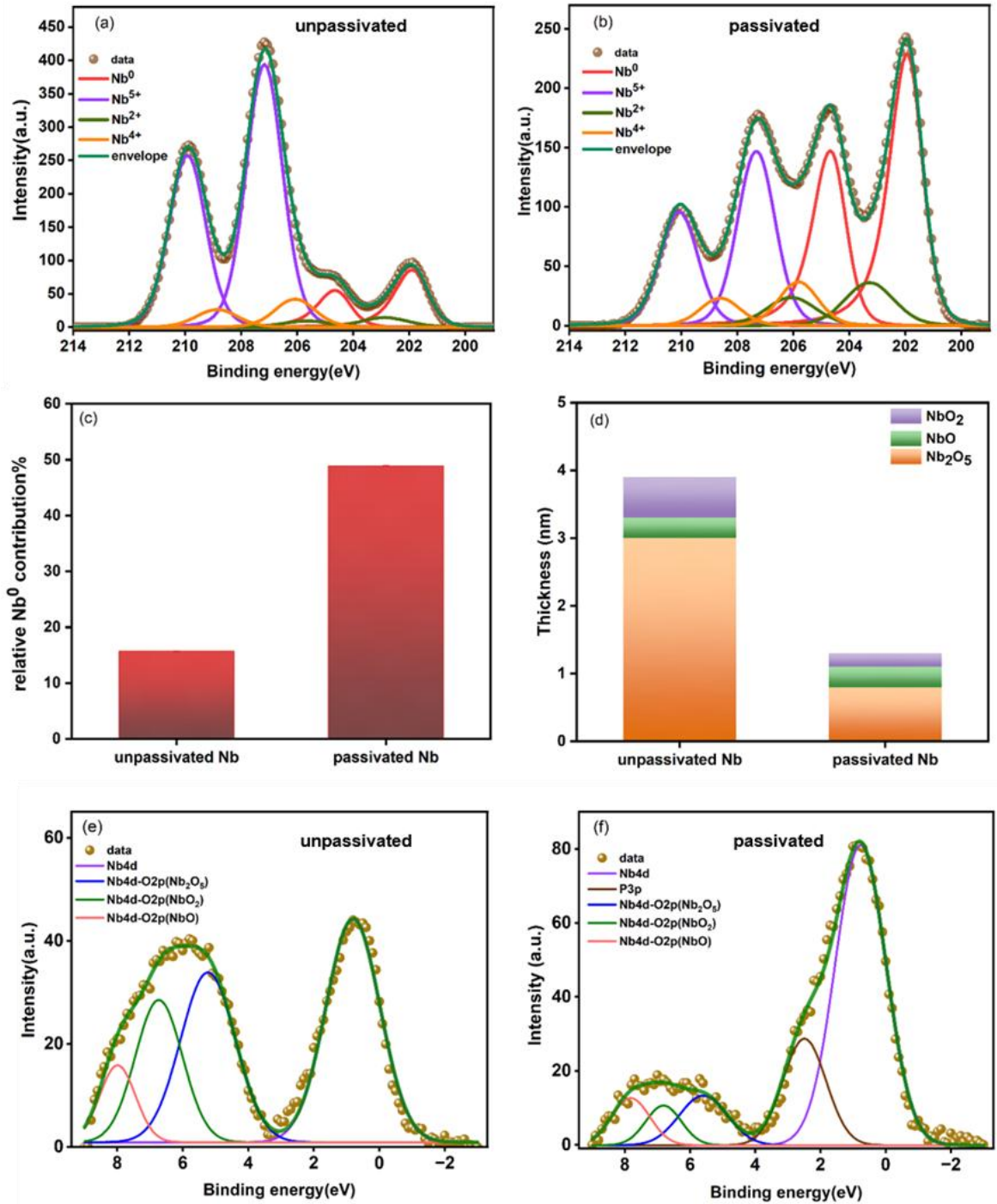

*Figure 10*: *(a) Background corrected Nb 3d XPS spectra for un-passivated Nb, aged for 6 days. (b) Background corrected Nb 3d XPS spectra for SAM-passivated Nb, aged for 6 days. Different oxidation states for Nb ($Nb^0$: elemental Nb, $Nb^{+2}$, $Nb^{+4}$ and $Nb^{+5}$) with spin-orbit coupled doublet ($3d_{5/2}$ and $3d_{3/2}$) are observed. (c) Relative $Nb^0$ (elemental Nb) contribution in un-passivated and passivated Nb, aged for 6 days; a clear damping of the signal is evident for un-passivated Nb. (d) The thicknesses of different oxides as extracted from the Nb 3d XPS spectra. For each thickness value, a relative error of ± 10% can be estimated. (e) Background corrected XPS - valence band spectra for an un-passivated Nb thin film, aged for 6 days. (f) XPS - valence band spectra for a passivated Nb thin film aged for 6 days. Valence band spectra indicate the different states attributed to Nb 4d and different hybridized (Nb4d-O2p) oxides states along with P 3p states for passivated films.*

## 3. Conclusion

In summary, the growth of highly ordered phosphonic acid SAMs on BOE-etched Nb for the passivation of superconducting resonators was achieved. Our study concludes that the loss components in un-passivated and passivated resonators are notably different in the higher critical photon number regime. In un-passivated resonators, this loss channel can be directly assigned to (regrown) $Nb_2O_5$, while the respective dominant loss channel in passivated resonators is ascribed to the passivating SAM itself, with an extracted $\tilde{\delta}_{SAM} \approx 5 \times 10^{-7}$. The SAM passivation provides a remarkable temporal stability to the Nb resonators in terms of $Q_i$ and resonance frequencies, when compared to the un-passivated resonators. This largely improved stability is attributed to an enhanced hydrophobicity of the surface and the ability of the SAM to restrict the regrowth of oxide, which thereby hinders the further accumulation of TLS defects. Still however, the $Q_i$ for passivated resonators remains lower in absolute numbers compared to the one of the un-passivated reference; this can possibly be improved with alternative SAM growth methods or by using different molecular coupling groups and backbones, which may also further reduce intrinsic losses for the passivated resonators. In this context, future work will also focus on a more detailed quantification of the remaining thin oxide at the interface to the SAM, and on strategies to completely avoid it.

## 4. Experimental Section/Methods

### Nb thin-film deposition and CPW resonator fabrication

150 nm thick Nb thin films (Fig. S1) were deposited on a high resistivity ($\rho > 10k\Omega.cm$) Si(100) substrate using an ultra-high vacuum sputtering tool (PLASSYS MEB550 S4-I) with a base pressure typically maintained below $1 \times 10^{-8}$ mbar. $\lambda/4$ co-planar waveguide resonators of different resonant frequencies in the range of 4-8GHz in hanging geometry were fabricated using photolithography and reactive ion etching (Supplementary Information: Nb thin-film deposition and CPW resonator fabrication).

### SAM growth and characterization

**Growth**

We have grown organophosphonate SAMs onto two different films: (i) planar Nb films on silicon substrates and (ii) Nb films on silicon substrates structured into CPW resonators. On both films, decylphosphonic acid ($C_{10}H_{23}O_3P$) molecules, with concentration 1mM in anhydrous toluene solution, were used for the SAM growth. (Supplementary Information: Self-assembled monolayer growth and characterization).

**Characterization**

The grown SAMs were characterized using contact angle (CA) goniometry, ellipsometry, X-ray photoelectron spectroscopy (XPS), Fourier-transform infrared (FTIR) spectroscopy and atomic force microscopy (AFM).

### Contact angle

To measure surface wettability and stability over time (during air exposure), we used a Dataphysics OCA15 water contact angle goniometry setup. Here, a 2 μL droplet of DI-water was casted on the probed surface using a precision dosing system. Thereafter, the angle between the surface and the water droplet meniscus (contact angle, CA) was measured using the Dataphysics software applied to photographs recorded immediately after the water droplet had attached to the sample surface. The contact angle values given in this work correspond to average values determined from multiple droplets and corresponding measurements performed on at least 5 different spots for each sample.

*X-ray photoelectron spectroscopy*

X-ray photoelectron spectroscopy (XPS) measurements were carried out in a homebuilt setup (parts from SPECS Surface Nano Analysis) under ultra-high vacuum (~6x10$^{-9}$ mbar). The setup is equipped with an XR 50 x-ray source and a Phoibos 100 hemispherical electron analyzer. A magnesium anode was utilized, generating X-ray photons of the energy of the Mg K$\alpha$ edge (E = 1253.6eV). The x-ray source was operated at a voltage of 12.50 kV and an emission current of 20.0 mA. Photoelectron spectra were acquired with an energy resolution of 0.1 eV, pass energy 25 eV and a dwell time of 1 s, and they were averaged over two to four scan repetitions. XPS data were analyzed with CasaXPS (www.casaxps.com; charge referencing, background subtraction) and Originlab software. The spectra were charge-referenced with respect to the C 1s hydrocarbon peak at 285 eV.

*Ellipsometry*

Variable angle spectroscopic ellipsometry measurements were made with an J.A. Woolman Co. alpha-SE ellipsometer at two different incidence angles (75° and 65°) over a wavelength range of 400-900 nm. The thicknesses of the SAMs were estimated from ellipsometry by fitting the top layer with Cauchy's dispersion relation. For the aliphatic SAMs, an isotropic refractive index $n$ = 1.5 was assumed for thickness estimation [65,66].

*Fourier-transform infrared (FTIR) spectroscopy*

The structural order and quality of the SAMs was investigated by infrared spectroscopy. FTIR spectra in attenuated total reflection (ATR) absorption mode were recorded at room temperature and under 0.3 mbar vacuum with a Vertex 70v spectrometer (Bruker), in a broad wavenumber range of 400 - 4000 cm$^{-1}$, to obtain the molecular vibration fingerprint of the SAM. The spectrometer was equipped with a Globar light source, a KBr beam splitter, and a nitrogen-cooled mercury cadmium telluride (MCT) detector. For the ATR measurements, the sample was placed upside down against a Ge hemispherical ATR crystal module with a built-in pressure applicator.

*Atomic force microscopy*

To determine the surface morphology and validate the conformal growth of the SAMs, AFM images were recorded in tapping mode using a Bruker (Veeco) Dimension V-AFM. Specifically, the surface roughness values before and after SAM passivation of the Nb thin films were measured. A scan size of 5x5μm$^2$ was taken for each sample at three different spots. The images were processed and analyzed for roughness estimation using the Gwyddion 2.66 software.

**Microwave measurements**

We investigated the RF response of Nb resonators in transmission mode at ~10mK by performing frequency domain measurements at varying input powers for freshly prepared and aged resonators. The samples were measured in a Bluefors Bottom-Loading Fast Sample Exchange System (Supplementary figure S6). The measurements involved the recording of the transmission coefficient $S_{21}$ for the resonators over the frequency range 4-8 GHz. The detailed measurement setup and analysis is discussed in the supplementary information (resonator measurement).

**Data Availability Statement**

The data that support the findings of this study are available from the corresponding authors upon reasonable request.

# References

1. Wang, C. *et al.* Towards practical quantum computers: transmon qubit with a lifetime approaching 0.5 milliseconds. *npj Quantum Inf* **8,** 1–6; 10.1038/s41534-021-00510-2 (2022).

2. Somoroff, A. *et al.* Millisecond Coherence in a Superconducting Qubit. *Physical review letters* **130,** 267001; 10.1103/PhysRevLett.130.267001 (2023).

3. Clarke, J. & Wilhelm, F. K. Superconducting quantum bits. *Nature* **453,** 1031–1042; 10.1038/nature07128 (2008).

4. Morten Kjaergaard *et al.* Superconducting Qubits: Current State of Play. *Annu. Rev. Condens. Matter Phys.* **11,** 369–395 (2020).

5. Müller, C., Cole, J. H. & Lisenfeld, J. Towards understanding two-level-systems in amorphous solids: insights from quantum circuits. *Reports on progress in physics. Physical Society (Great Britain)* **82,** 124501; 10.1088/1361-6633/ab3a7e (2019).

6. Burnett, J., Bengtsson, A., Niepce, D. & Bylander, J. Noise and loss of superconducting aluminium resonators at single photon energies. *J. Phys.: Conf. Ser.* **969,** 12131; 10.1088/1742-6596/969/1/012131 (2018).

7. Richardson, C. J. K. *et al.* Fabrication artifacts and parallel loss channels in metamorphic epitaxial aluminum superconducting resonators. *Supercond. Sci. Technol.* **29,** 64003; 10.1088/0953-2048/29/6/064003 (2016).

8. Goetz, J. *et al.* Loss mechanisms in superconducting thin film microwave resonators. *Journal of Applied Physics* **119,** 15304; 10.1063/1.4939299 (2016).

9. Oliver, W. D. & Welander, P. B. Materials in superconducting quantum bits. *MRS Bull.* **38,** 816–825; 10.1557/mrs.2013.229 (2013).

10. Altoé, M. V. P. *et al.* Localization and Mitigation of Loss in Niobium Superconducting Circuits. *PRX Quantum* **3,** 20312; 10.1103/PRXQuantum.3.020312 (2022).

11. Gorgichuk, N., Junginger, T. & Sousa, R. d. Modelling dielectric loss in superconducting resonators: Evidence for interacting atomic two-level systems at the Nb/oxide interface. *Phys. Rev. Applied* **19**; 10.1103/PhysRevApplied.19.024006 (2023).

12. Verjauw, J. *et al.* Investigation of Microwave Loss Induced by Oxide Regrowth in High-Q Niobium Resonators. *Phys. Rev. Applied* **16,** 14018; 10.1103/PhysRevApplied.16.014018 (2021).

13. Karuppannan, S. K. *et al.* Improved Interface of Niobium Superconducting Resonator with Ruthenium as a Capping Layer. *ACS Appl. Electron. Mater.* **6,** 7372–7379; 10.1021/acsaelm.4c01268 (2024).

14. Bal, M. *et al.* Systematic improvements in transmon qubit coherence enabled by niobium surface encapsulation. *npj Quantum Inf* **10**; 10.1038/s41534-024-00840-x (2024).

15. Badia, A., Lennox, R. B. & Reven, L. A dynamic view of self-assembled monolayers. *Accounts of chemical research* **33,** 475–481; 10.1021/ar9702841 (2000).

16. Frank Schreiber. Structure and growth of self-assembling monolayers. *Progress in Surface Science* **65,** 151 (2000).

17. Abraham Ulman. Formation and Structure of Self-Assembled Monolayers. *Chemical reviews* **96,** 1533–1554 (1996).

18. Gooding, J. J., Mearns, F., Yang, W. & Liu, J. Self‐Assembled Monolayers into the 21 st Century: Recent Advances and Applications. *Electroanalysis* **15,** 81–96; 10.1002/elan.200390017 (2003).

19. Singh, M., Kaur, N. & Comini, E. The role of self-assembled monolayers in electronic devices. *J. Mater. Chem. C* **8,** 3938–3955; 10.1039/D0TC00388C (2020).

20. Th. Wink, S. J. van Zuilen, A. Bult and W. P. van Bennekom. Self-assembled Monolayers for Biosensors. *Analyst* **122,** 43R-53R (1997).

21. Taneichi, D., Haneda, R. & Aramaki, K. A novel modification of an alkanethiol selfassembled monolayer with alkylisocyanates to prepare protective films against copper corrosion. *Corrosion Science* **43,** 1589–1600 (2001).

22. Hutt, D. A. & Liu, C. Oxidation protection of copper surfaces using self-assembled monolayers of octadecanethiol. *Applied Surface Science* **252,** 400–411; 10.1016/j.apsusc.2005.01.019 (2005).

23. Wang, S. *et al.* The Role of Self-Assembled Monolayers in the Surface Modification and Interfacial Contact of Copper Fillers in Electrically Conductive Adhesives. *ACS applied materials & interfaces* **16,** 1846–1860; 10.1021/acsami.3c14900 (2024).

24. Pujari, S. P., Scheres, L., Marcelis, A. T. M. & Zuilhof, H. Covalent surface modification of oxide surfaces. *Angewandte Chemie (International ed. in English)* **53,** 6322–6356; 10.1002/anie.201306709 (2014).

25. Love, J. C., Estroff, L. A., Kriebel, J. K., Nuzzo, R. G. & Whitesides, G. M. Self-assembled monolayers of thiolates on metals as a form of nanotechnology. *Chemical reviews* **105,** 1103–1169; 10.1021/cr0300789 (2005).

26. Ikegami, A., Suda, M., Watanabe, T. & Einaga, Y. Reversible optical manipulation of superconductivity. *Angewandte Chemie (International ed. in English)* **49,** 372–374; 10.1002/anie.200904548 (2010).

27. H. M. MCCONNELL, F. R. GAMBLE, B. M. HOFFMAN. Interactions betwqeen Interactions between superconductors and organic molecules. *Proceedings of the National Academy of Sciences* **57,** 1131 (1967).

28. Alghadeer, M. *et al.* Surface Passivation of Niobium Superconducting Quantum Circuits Using Self-Assembled Monolayers. *ACS applied materials & interfaces* **15,** 2319–2328; 10.1021/acsami.2c15667 (2023).

29. Alghadeer, M. *et al.* Mitigating coherent loss in superconducting circuits using molecular self-assembled monolayers. *Scientific reports* **14,** 27340; 10.1038/s41598-024-77227-7 (2024).

30. Cattani-Scholz, A. Functional Organophosphonate Interfaces for Nanotechnology: A Review. *ACS applied materials & interfaces* **9,** 25643–25655; 10.1021/acsami.7b04382 (2017).

31. Wei Gao, Lucy Dickinson, Christina Grozinger, Frederick G. Morin, andLinda Reven. Self-Assembled Monolayers of Alkylphosphonic Acids onMetal Oxides. *Langmuir* **12,** 6249–6435 (1996).

32. Cattani-Scholz, A. *et al.* Organophosphonate-based PNA-functionalization of silicon nanowires for label-free DNA detection. *ACS nano* **2,** 1653–1660; 10.1021/nn800136e (2008).

33. Ma, H., Acton, O., Hutchins, D. O., Cernetic, N. & Jen, A. K.-Y. Multifunctional phosphonic acid self-assembled monolayers on metal oxides as dielectrics, interface modification layers and semiconductors for low-voltage high-performance organic field-effect transistors. *Physical chemistry chemical physics : PCCP* **14,** 14110–14126; 10.1039/c2cp41557g (2012).

34. Dlugosch, J. M. *et al.* Conductance Switching in Liquid Crystal-Inspired Self-Assembled Monolayer Junctions. *ACS applied materials & interfaces* **14,** 31044–31053; 10.1021/acsami.2c05264 (2022).

35. Vilan, A., Aswal, D. & Cahen, D. Large-Area, Ensemble Molecular Electronics: Motivation and Challenges. *Chemical reviews* **117,** 4248–4286; 10.1021/acs.chemrev.6b00595 (2017).

36. Anshuma Pathak. Structural and electronic properties of organophosphonate monolayers on alumina and silicon oxide. PhD thesis. Technischen Universität Braunshweig, 2016.

37. Bauer, T. *et al.* Phosphonate- and carboxylate-based self-assembled monolayers for organic devices: a theoretical study of surface binding on aluminum oxide with experimental support. *ACS applied materials & interfaces* **5,** 6073–6080; 10.1021/am4008374 (2013).

38. Shaheen, A. *et al.* Characterization of Self-Assembled Monolayers on a Ruthenium Surface. *Langmuir : the ACS journal of surfaces and colloids* **33,** 6419–6426; 10.1021/acs.langmuir.7b01068 (2017).

39. Pathak, A. *et al.* Disorder-derived, strong tunneling attenuation in bis-phosphonate monolayers. *Journal of physics. Condensed matter : an Institute of Physics journal* **28,** 94008; 10.1088/0953-8984/28/9/094008 (2016).

40. Pathak, A. *et al.* Nanocylindrical confinement imparts highest structural order in molecular self-assembly of organophosphonates on aluminum oxide. *Nanoscale* **9,** 6291–6295; 10.1039/c7nr02420g (2017).

41. Tai, T., Cai, J. & Anlage, S. M. Anomalous Loss Reduction Below Two‐Level System Saturation in Aluminum Superconducting Resonators. *Adv Quantum Tech* **7,** 2200145; 10.1002/qute.202200145 (2024).

42. Kirsh, N., Svetitsky, E., Burin, A. L., Schechter, M. & Katz, N. Revealing the nonlinear response of a tunneling two-level system ensemble using coupled modes. *Phys. Rev. Materials* **1,** 12601; 10.1103/PhysRevMaterials.1.012601 (2017).

43. Macha, P. *et al.* Losses in coplanar waveguide resonators at millikelvin temperatures. *Applied Physics Letters* **96,** 62503; 10.1063/1.3309754 (2010).

44. Romanenko, A. & Schuster, D. I. Understanding Quality Factor Degradation in Superconducting Niobium Cavities at Low Microwave Field Amplitudes. *Phys. Rev. Lett.* **119,** 264801; 10.1103/PhysRevLett.119.264801 (2017).

45. Wang, H. *et al.* Improving the coherence time of superconducting coplanar resonators. *Applied Physics Letters* **95,** 233508; 10.1063/1.3273372 (2009).

46. Faoro, L. & Ioffe, L. B. Internal loss of superconducting resonators induced by interacting two-level systems. *Physical review letters* **109,** 157005; 10.1103/PhysRevLett.109.157005 (2012).

47. Graaf, S. E. de *et al.* Suppression of low-frequency charge noise in superconducting resonators by surface spin desorption. *Nature communications* **9,** 1143; 10.1038/s41467-018-03577-2 (2018).

48. Müller, C., Lisenfeld, J., Shnirman, A. & Poletto, S. Interacting two-level defects as sources of fluctuating high-frequency noise in superconducting circuits. *Phys. Rev. B* **92,** 35442; 10.1103/PhysRevB.92.035442 (2015).

49. Wenner, J. *et al.* Surface loss simulations of superconducting coplanar waveguide resonators. *Applied Physics Letters* **99**; 10.1063/1.3637047 (2011).

50. Kopas, C. J. *et al.* Low microwave loss in deposited Si and Ge thin-film dielectrics at single-photon power and low temperatures. *AIP Advances* **11,** 95007; 10.1063/5.0041525 (2021).

51. Li, D. *et al.* Improvements in Silicon Oxide Dielectric Loss for Superconducting Microwave Detector Circuits. *IEEE Trans. Appl. Supercond.* **23,** 1501204; 10.1109/TASC.2013.2242951 (2013).

52. Mittal, S. *et al.* Annealing reduces Si3N4 microwave-frequency dielectric loss in superconducting resonators. *Phys. Rev. Applied* **21,** 54044; 10.1103/PhysRevApplied.21.054044 (2024).

53. Krupka, J. Measurements of the Complex Permittivity of Low Loss Polymers at Frequency Range From 5 GHz to 50 GHz. *IEEE Microw. Wireless Compon. Lett.* **26,** 464–466; 10.1109/LMWC.2016.2562640 (2016).

54. Bruno, A. *et al.* Reducing intrinsic loss in superconducting resonators by surface treatment and deep etching of silicon substrates. *Applied Physics Letters* **106,** 182601; 10.1063/1.4919761 (2015).

55. Kumar, S. *et al.* Temperature dependence of the frequency and noise of superconducting coplanar waveguide resonators. *Applied Physics Letters* **92,** 123503; 10.1063/1.2894584 (2008).

56. Simpson, J. T., Hunter, S. R. & Aytug, T. Superhydrophobic materials and coatings: a review. *Reports on progress in physics. Physical Society (Great Britain)* **78,** 86501; 10.1088/0034-4885/78/8/086501 (2015).

57. Buabthong, P., Becerra Stasiewicz, N., Mitrovic, S. & Lewis, N. S. Vanadium, niobium and tantalum by XPS. *Surface Science Spectra* **24,** 24001; 10.1116/1.4998018 (2017).

58. Grundner, M. & Halbritter, J. XPS and AES studies on oxide growth and oxide coatings on niobium. *Journal of Applied Physics* **51,** 397–405; 10.1063/1.327386 (1980).

59. Prudnikava, A. *et al.* Systematic study of niobium thermal treatments for superconducting radio frequency cavities employing x-ray photoelectron spectroscopy. *Supercond. Sci. Technol.* **35,** 65019; 10.1088/1361-6668/ac6a85 (2022).

60. Cumpson, P. J. & Seah, M. P. Elastic Scattering Corrections in AES and XPS. II. Estimating Attenuation Lengths and Conditions Required for their Valid Use in Overlayer/Substrate Experiments. *Surf. Interface Anal.* **25,** 430–446; 10.1002/(SICI)1096-9918(199706)25:6<430::AID-SIA254>3.0.CO;2-7 (1997).

61. Kuznetsov, M. V., Razinkin, A. S. & Shalaeva, E. V. Photoelectron spectroscopy and diffraction of surface nanoscale Nb)/Nb(110) structures. *Journal of Structural Chemistry* **50,** 514–521 (2009).

62. Sanjinés, R., Benkahoul, M., Papagno, M., Lévy, F. & Music, D. Electronic structure of Nb2N and NbN thin films. *Journal of Applied Physics* **99,** 44911; 10.1063/1.2173039 (2006).

63. Elsener, B., Atzei, D., Krolikowski, A. & Rossi, A. Effect of phosphorus concentration on the electronic structure of nanocrystalline electrodeposited Ni–P alloys: an XPS and XAES investigation. *Surface & Interface Analysis* **40,** 919–926; 10.1002/sia.2802 (2008).

64. Nancy B. Goodman, L. Ley & and D. W. Bullett. Valence-band structures of phosphorus allotropes. *Phys. Rev. B* **27,** 7440–7450 (1983).

65. Stephen R. Wasserman *et al.* The structure of self-assembled monolayers of alkylsiloxanes on silicon: a comparison of results from ellipsometry and low-angle x-ray reflectivity. *J. Am. Chem. Soc.* **111,** 5852–5861 (1989).

66. Atul N. Parikh, David L. Allara, Issam Ben Azouz & Francis Rondelez. An Intrinsic Relationship between Molecular Structure in Self-Assembled n-Alkylsiloxane Monolayers and Deposition Temperature. *J. Phys. Chem.* **98,** 7577–7590 (1994).

**Acknowledgements**

We acknowledge funding through the Munich Quantum Valley (MQV) project by the Free State of Bavaria, Germany. The Walter Schottky Institute of TU Munich (Prof. Sharp, Prof. Stutzmann) is gratefully acknowledged for providing access to their XPS facility. We thank the staff of the ZEITlab shared facilities of TU Munich for expert technical support. Franz Haslbeck from the Walther Meiβner Institute as well as Johannes Weber, Daniela Zahn, Simon Lang and Alexandra Schewski from the Fraunhofer EMFT are acknowledged for scientific discussion and support.

**Author contributions**

M.T and H.G conceived the idea. H.G., L.K., R.P. and N.B. designed the experiments. M.T. and S.F. provided the experimental facilities and manage the project. H.G., L.K., R.P., N.B., M.S., B.S. and M.K. carried out the experiments. H.G., R.P., L.K. and M.T. analyzed the data and interpreted this study. H.G. wrote the manuscript. R.P., M.T., L.K., B.S. and M.K. revised the manuscript. All authors discussed the results and commented on the manuscript.

**Competing interests**

The authors declare no competing interests.

# Supplementary Information

# High temporal stability of niobium superconducting resonators by surface passivation with organophosphonate self-assembled monolayers

*Harsh Gupta\*[1], Rui Pereira[2], Leon Koch[3,4], Niklas Bruckmoser[3,4], Moritz Singer[1], Benedikt Schoof[1], Manuel Kompatscher[1], Stefan Filipp[3,4], Marc Tornow\*[1,2]*

[1]*TUM School of Computation, Information and Technology, Department of Electrical Engineering, Technical University of Munich, 85748 Garching, Germany*

E-mail\*: harsh.gupta@tum.de, tornow@tum.de

[2] *Fraunhofer Institute for Electronic Microsystems and Solid-State Technologies, 80686 Munich, Germany*

[3]*Walther-Meißner-Institut, Bayerische Akademie der Wissenschaften, 85748 Garching, Germany*

[4] *TUM School of Natural Sciences, Department of Physics, Technical University of Munich, 85748 Garching, Germany*

**Nb thin-film deposition and CPW resonator fabrication**

The fabrication process begins with cleaning a float-zone silicon substrate featuring a resistivity above 10kΩ.cm. To clean the silicon, we first immerse the substrate in a piranha solution (a 3:1 mixture of sulfuric acid at 30% concentration and hydrogen peroxide) *[caution: corrosive and strong oxidizer; special care/training is mandatory]* at 80°C for 10min. This strongly oxidizing solution effectively removes organic residues from the substrate surface. Following this cleaning step, a thin silicon oxide layer forms on the substrate. To remove this oxide, we immerse the substrate in a buffered oxide etch (BOE) solution (7:1) (HF: $NH_4F$) for 30sec *[caution: hydrofluoric acid (HF(aq)) is very hazardous to health; special care/training is mandatory]*. The substrate is then rinsed with deionized water and transferred to an ultra-high vacuum (UHV) sputter chamber (PLASSYS MEB550 S4-I) with a base pressure typically maintained below $5 \times 10^{-9}$ mbar. In the UHV chamber, we deposit 150 nm thick niobium thin film onto the substrate via DC magnetron sputtering. After sputtering, the niobium-coated substrate is removed from the chamber, and an optical resist layer (AZ MiR 701) is spun onto the surface. The resist is exposed using maskless lithography and developed in AZ 726 MIF, a TMAH-based developer. The substrate is then transferred to a reactive ion etching (RIE) system from Oxford Instruments, where the niobium thin film is patterned using an $SF_6$ based plasma etch. Next, the photoresist mask is removed, followed by a final rinse with deionized water. Co-planar waveguide resonators in hanger geometry (up to 13 different resonant frequencies) were fabricated on each chip.

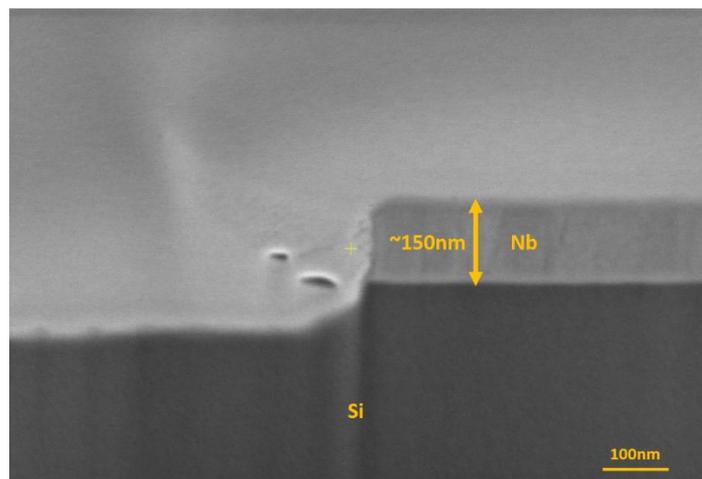

*Figure S1* Focused ion beam cut cross-sectional SEM image of a Nb resonator thin-film; the tilt-corrected Nb film thickness is measured to be 150nm, as indicated.

**Self-assembled monolayer growth and characterization**

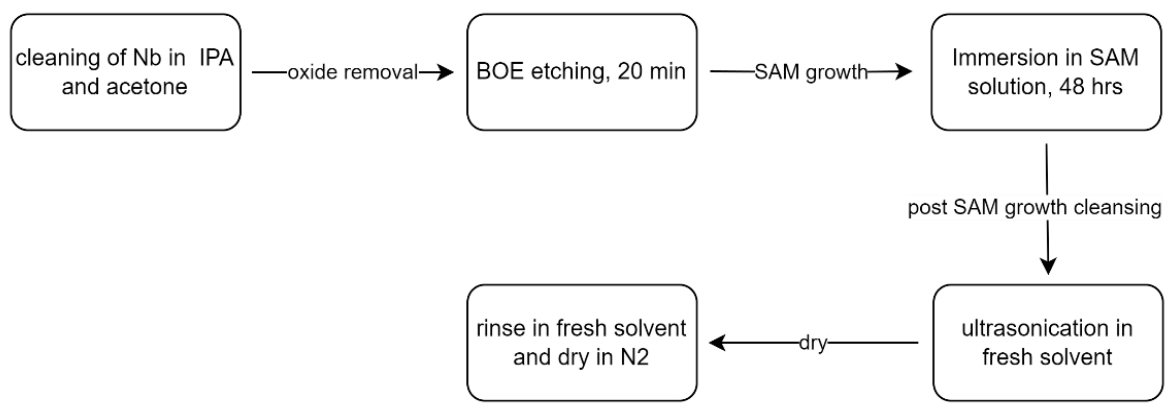

*Figure S2* Flow diagram for organophosphonate SAM growth on BOE-etched Nb.

*Growth recipe:* The following recipe is employed for the growth of phosphonic acid SAMs on Nb films and resonators. First, a SAM-molecules stock solution of concentration 20mM was prepared with decylphosphonic acid molecules (from Sigma-Aldrich) in anhydrous toluene (from Sigma-Aldrich) at room temperature, which was further used to prepare a solution of concentration 1mM by dilution, to be used in the present work. Nb thin films and resonators were first cleaned with acetone followed by propanol and DI water using ultrasonication for 3 min each, and then etched in BOE (7:1, HF: $NH_4F$ = 12.5: 87.5%) solution for 20min to remove the native oxide. Subsequently, the etched substrates were promptly (within 2 mins) immersed in the SAM solution, where SAMs were then grown from the liquid phase on the Nb surface. For this purpose, they were kept in solution for 48 hrs and remained undisturbed during this time to facilitate SAM growth. After SAM growth, the samples were rinsed in fresh, clean toluene solvent and dried in a nitrogen flow.

*Ellipsometry:* The thickness of the SAMs and the oxide was estimated using ellipsometry. Spectroscopic ellipsometry data was recorded over a wavelength range 380-900 nm using a JA Woollam ellipsometer at room temperature for 2 samples with 3 different spots on each sample. To extract the thickness of the SAM and the underlying oxide we employed a three-layer model for fitting of the ellipsometry data. The Nb layer was fitted using a b-spline model, whereas the

NbO$_x$ and the SAM layers were fitted using Cauchy's model; refractive index values of 2.2 and 1.5 were used, respectively, for the NbO$_x$ and SAM Cauchy layers for thickness estimation[1,2]. Data was fitted with mean-squared error of <10. Extracted thicknesses are given in table S1.

*Table S1 Thicknesses of oxide and SAM as extracted from ellipsometry data for un-passivated and passivated Nb films.*

| Layer | NbOx | SAM |
| --- | --- | --- |
| Thickness (un-passivated) | 3.4±0.1nm | -- |
| Thickness (passivated) | 1.0±0.2nm | 1.2±0.2nm |

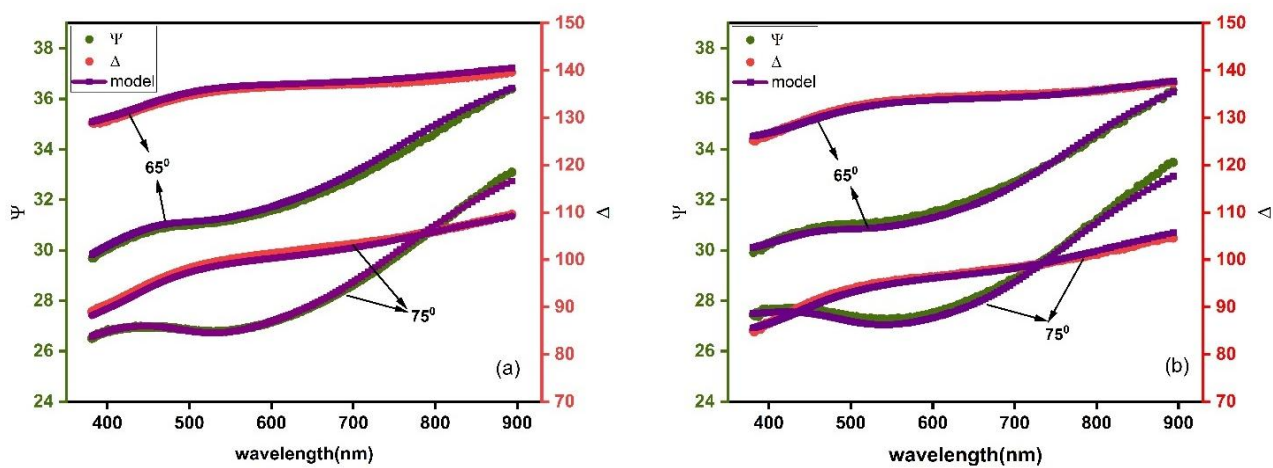

*Figure S3 Spectroscopic ellipsometry data recorded at two different incidence angles of 65° and 75° for (a), a passivated Nb thin film and (b), an un-passivated Nb thin film. A three-layer model, one b-spline and two Cauchy layers for the passivated, and a two-layer model, one b-spline and one Cauchy layer for the un-passivated sample, respectively, were fitted to Ψ and Δ to extract the thicknesses of layers.*

**X-ray photoelectron spectroscopy (XPS)**

First, XPS survey spectra was recorded for all samples, all the constituent elements were then identified from survey spectra, in our case Nb (3d around 200eV, 3p around 370eV), O (1s around 532eV), C (1s around 285eV), P (2p around 134eV). Thereafter, a fine resolution spectrum for all elements was recorded. For all the spectra, a charge referencing correction was performed considering the C1s hydrocarbon peak at 285eV. The high-resolution P-2p spectra was collected for both un-passivated and passivated Nb films. An intense P 2p signal (centered at ~134eV) is observed on passivated Nb whereas no signal is detected for un-passivated films (figure S4); this affirms the presence of the phosphonic acid molecule SAM on the passivated Nb.

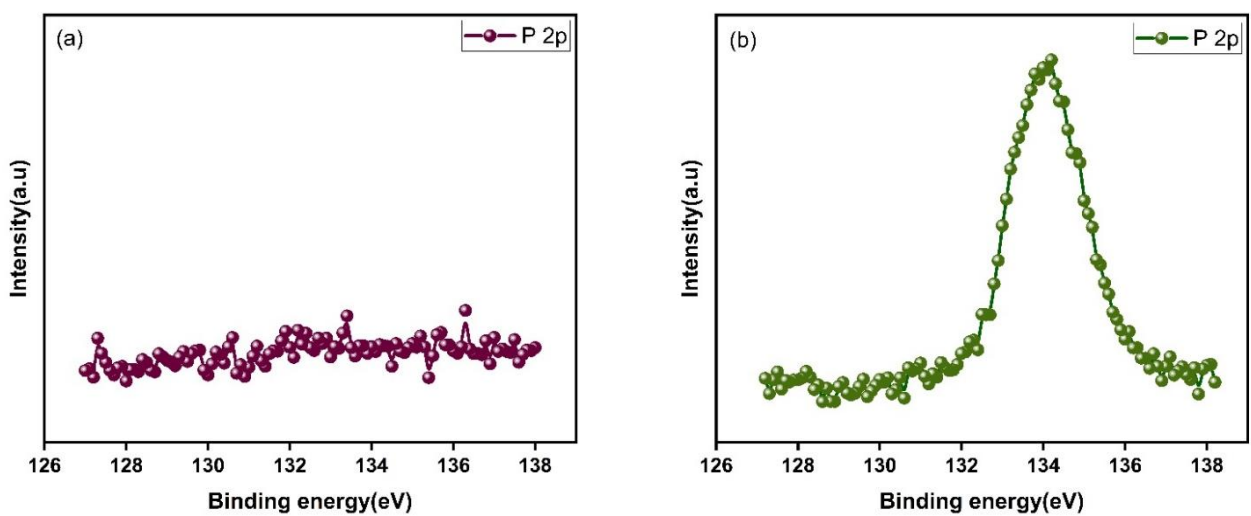

*Figure S4* High resolution P 2p spectra for (a) un-passivated Nb, where no phosphorus is detected, and (b) passivated Nb, showing the presence of phosphorus in the phosphonic acid molecules constituting the SAM.

Fitting routine for Nb spectrum: For detailed analysis of the Nb surface chemistry, we considered the Nd-3d spectrum for both un-passivated and passivated Nb, aged for 6 days. We subtracted the Shirley background from the spectra in the binding energy range of interest (200 213eV)[3]. Different components were fitted with a LA (*a,b,c*) line-shape which is a superset of the Voigt function, in CASA XPS, where *a, b* defines the spread of the tail on either side of the Lorentzian component and the parameter *c* specifies the width of the Gaussian used to convolute the Lorentzian curve. Different oxidation states for Nb were identified; for the fitting routine, spin-orbit doublets ($3d_{5/2, 3/2}$) were considered for all suboxides with fixed energy separation of 2.7eV and the defined area ratio of 3:2 for $d_{5/2}$:$d_{3/2}$ peaks, for both passivated and un-passivated Nb. The $Nb^0$ oxidation state corresponding to metallic Nb was fitted with an asymmetric peak, as for metals unfilled one-electron levels (the conduction band), can accept electrons that have undergone shake-up type processes following ejection of the initial core electron [4]. As a result, rather than displaying a distinct structure typical of shake-up satellites, a tail appears on the higher binding energy side of the main peak, this feature can be manifested in an asymmetric peak shape [5,6]. Whereas, other oxidation states ($Nb^{+2}$, $Nb^{+4}$, $Nb^{+5}$) corresponding to $NbO_x$ show symmetric peaks and are fitted with the same FWHM for individual spin-orbit doublets for different suboxides. The implementation of the asymmetric Lorentzian lineshape in CasaXPS includes a parameter specifying the width of the Gaussian used to convolute the Lorentzian curve, namely LA(α, β, m), where m is an integer defining the width of the Gaussian (see Table S2).

*Valence Band spectra*: A high-resolution valence band (VB) spectrum in the energy range 9eV to -3eV was recorded for un-passivated and passivated Nb thin films. Background correction similar to core states was performed for valence band spectra; for fitting, Gaussian peaks were used, same FWHM were used for Nb 4d and different oxides, respectively, with 5% variation. VB spectra was fitted with a LA (1.53, 2.43) line shape for different components (Nb 4d, Nb 4d-O2p and P 3p).

*Table S2 Peak positions, line shapes and FWHM for the different components of the Nb 3d XPS spectra for unpassivated Nb thin films. If in the parameter set only two numbers are given, then these correspond to a and b. Similar parameters were used for passivated Nb thin films within 5% deviation.*

|  | Position(eV) | Line shape ($\alpha,\beta,m$) | FWHM |
| --- | --- | --- | --- |
| $Nb^0$ | 201.9 | LA (1.30, 2.44, 69) | 1.28 |
| $Nb^0$ | 204.6 | LA (1.30, 2.44, 69) | 1.31 |
| $Nb^{+5}$ | 207.2 | LA (1.53, 2.43) | 1.53 |
| $Nb^{+5}$ | 209.9 | LA (1.53, 2.43) | 1.55 |
| $Nb^{+4}$ | 206.1 | LA (1.53, 2.43) | 1.70 |
| $Nb^{+4}$ | 208.8 | LA (1.53, 2.43) | 1.74 |
| $Nb^{+2}$ | 202.9 | LA (1.53, 2.43) | 1.94 |
| $Nb^{+2}$ | 205.6 | LA (1.53, 2.43) | 1.98 |

*XPS Straight Line - Attenuation (SLA) Length Model*: We take the simple case of an overlayer A (different suboxides) on a substrate S (Nb in present case), each being amorphous or polycrystalline. In the SLA, it is assumed that electrons follow straight-line paths from creation to emission (i.e., no elastic scattering). In such case for an overlayer A (oxide of substrate S), the thickness (d) can be estimated using expression equation S1 [7,8].

$$d \propto \lambda_i \ln\left[1 + \frac{\left(\frac{I_A}{S_A}\right)}{\left(\frac{I_S}{S_S}\right)}\right] \qquad (S1)$$

Where, $\lambda_i$ is the inelastic mean free path of electrons, $I_A$ and $I_S$ are the intensities of electrons and $S_A$ and $S_S$ are the relative sensitivity factors, each for the overlayer A and the substrate S, respectively. For the estimation of thicknesses, $\lambda_i$ and sensitivity factors are extracted from the SESSA and CasaXPS software. $\lambda_i$ was calculated to be 1.9 nm for $Nb_2O_5$, the same $\lambda_i$ was assumed for all suboxides. In the case of passivated samples an organic overlayer (thickness 1.2 nm) correction was considered to account for the damping of the Nb signal due to the SAM layer.

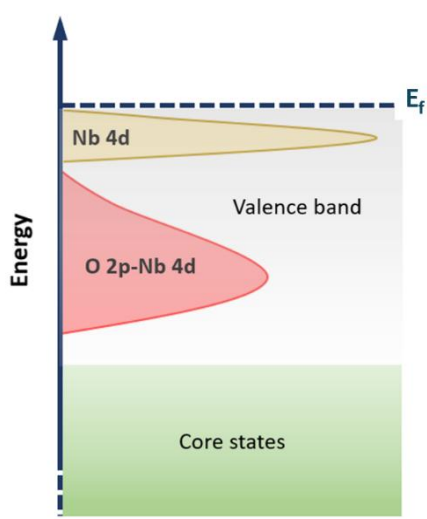

**Figure S5** Schematic energy band diagram for XPS spectra showing the specific valence band states probed; the core states participate in the main XPS spectra (e.g., Nb3d and P2p).

**Resonator measurements**

Frequency domain transmission spectra ($S_{21}$) in the microwave regime were recorded using a Keysight Vector Network Analyzer (VNA), with the sample mounted in a Bluefors Bottom-Loading Fast Sample Exchange dilution refrigerator operating at ~10mK. The resonator sample is housed in a copper two-port box. The RF signal was routed through wire bonds from SMA connectors located on the two-port copper housing towards the coplanar waveguide resonators. Further, to provide grounding for the resonator chip, the sample's ground plane was wire-bonded to the copper box. On the input port, -60 dB of attenuation across the various temperature stages and on the output line, a HEMT amplifier at 4K and a amplifier operating at 10 mK were installed. Following a band pass filter, the input signal is fed into the resonator chip under test. The output line has a band pass filter and circulators to diminish noise and to avoid signal backscattering. Subsequently, the resulting $S_{21}$ response goes back to the VNA and is analyzed further.

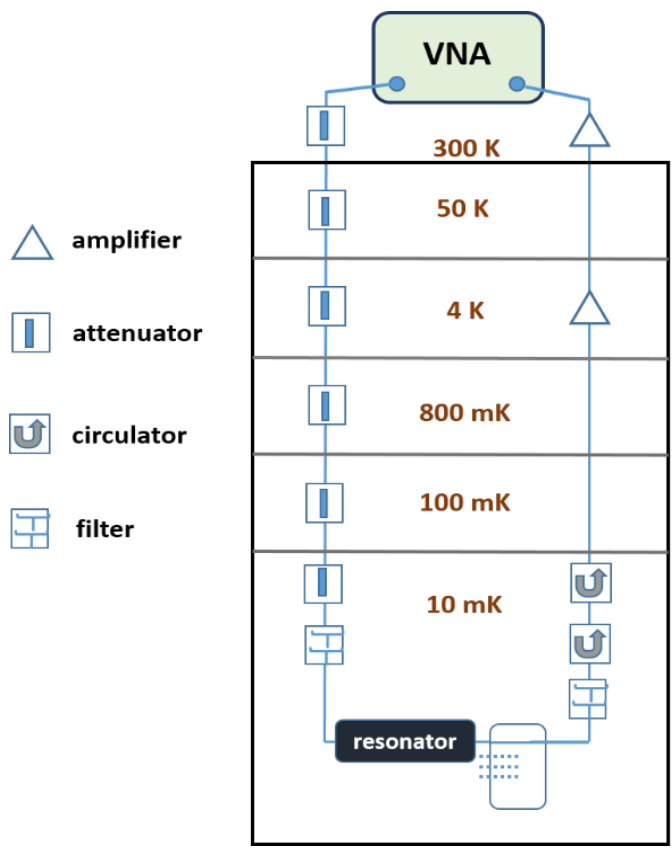

*Figure S6* Schematic setup for low temperature, microwave regime resonator measurements. The input signal propagates via the left (output) port of the VNA into the cryostat, in which the power is attenuated by up to - 60 dB. After running through the resonator device under test, the transmission output signal is amplified by low-noise HEMT amplifiers and fed back into the right (input) port of the VNA for analysis.

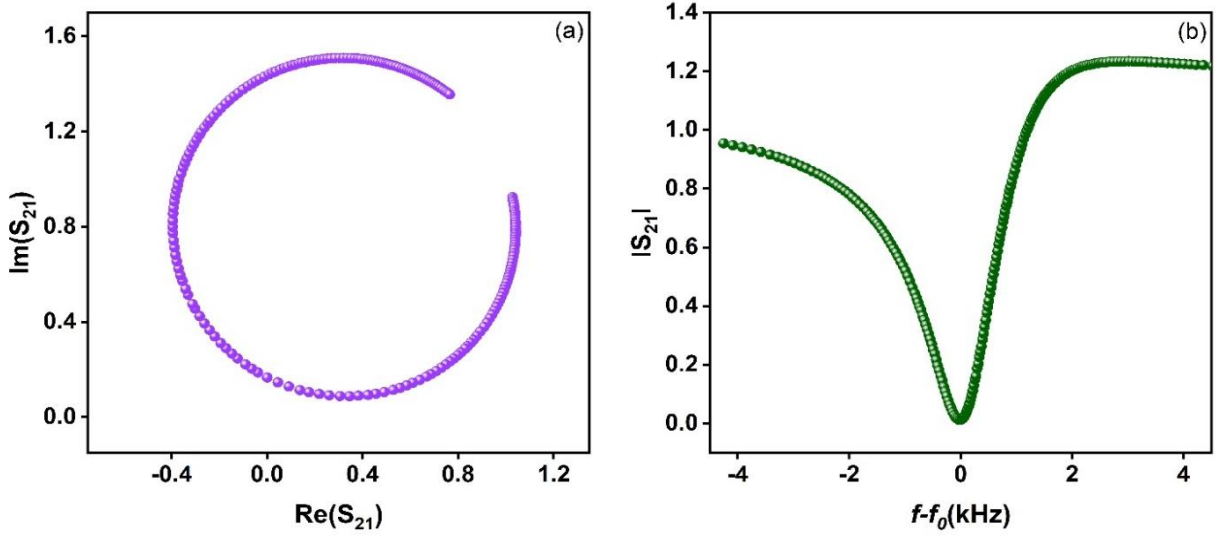

*Figure S7 Exemplary resonator response traces measured in transmission mode in the few GHz regime (a) Circular response representation in the complex $S_{21}$ plane, with a diameter $|Q_l|/|Q_c|$. (b) Lorentzian resonance response in the frequency domain.*

*Analysis:*
The values for the loaded quality factors $Q_l$ and the coupling quality factors $Q_c$ were extracted, by fitting the complex-valued frequency domain $S_{21}$ data with a circle fit according to [9–11]:

$$S_{21}(f) = ae^{i\alpha}e^{-2\pi i f\tau}\left[1 - \frac{(Q_l/|Q_c|)e^{i\varphi}}{1+2iQ_l(\frac{f}{f_r}-1)}\right] \quad (S2)$$

Here, $f$ is the probe frequency, $f_r$ is the resonator frequency, and $\varphi$ is the argument of the complex-valued coupling quality factor, which considers a possible impedance mismatch. To take the damping of the cables into consideration, the amplitude is modified by the factor $a$, and an initial offset of $\alpha$ is introduced in the phase along with a delay $\tau$ due to the finite length of the measurement cables. The microwave power inserted into the resonator circuit can be translated into a mean photon number $<n>$, which can be estimated from the following relation, with $P_{in}$ being the input microwave power [12,13]:

$$\langle n \rangle = \frac{Q_l^2}{\pi h f_r^2 Q_c}\frac{Z_0}{Z_r}P_{in} \quad (S3)$$

where $Z_o$ is the characteristic impedance of the microwave environment to which the resonator is coupled and $Z_r$ is the characteristic impedance of the resonator.

**Frequency dispersion with aging**

The aging of resonators, which results in the accumulation of TLS losses, not only affects the Q-factor but also significantly impacts their resonant frequencies. At constant probing temperature, the magnitude of relative change in frequencies is proportional to the TLS density as, $\frac{\Delta f_r}{f_r} \propto \delta_{TLS}$. It explains the greater relative shift of the resonance curve for un-passivated resonators as they accumulate higher TLS losses with aging. Below is an exemplary graph showing the shifts of the resonance curve of particular resonators (same resonance frequency, both passivated and un-passivated) with aging.

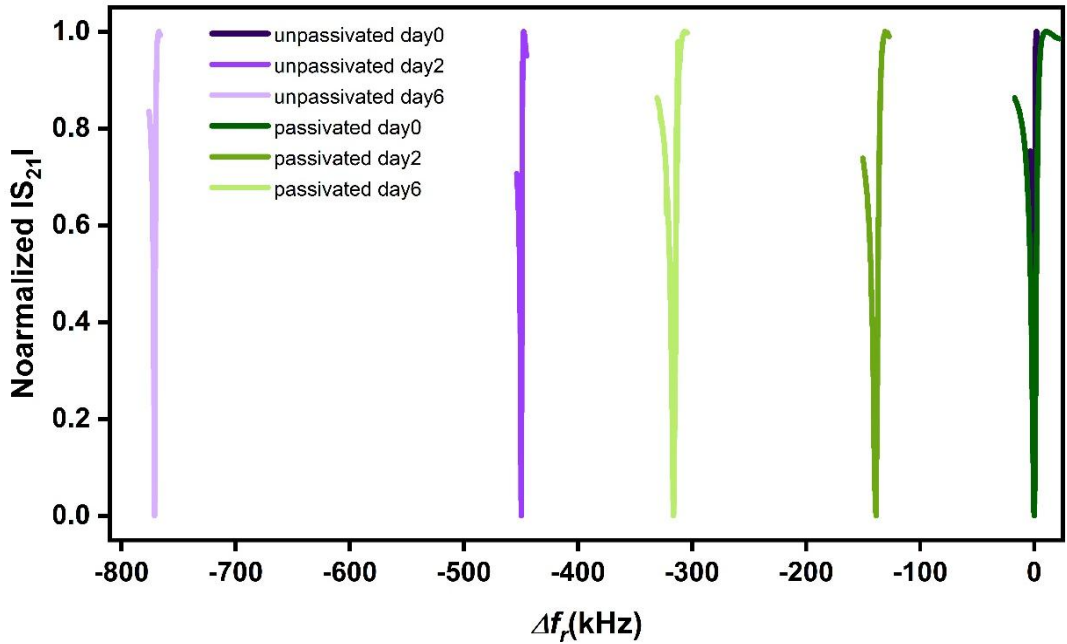

*Figure S8* *Shift in resonance frequency with aging, for an un-passivated and a passivated resonator, both with a day0 resonant frequency of 6.4GHz. The resonance curve for the un-passivated resonator shows a significantly larger shift in frequency with aging in comparison to the passivated resonator.*

**TLS loss model fitting procedure**

The $Q_i$-$n$ relations for the resonators of different frequencies on passivated and un-passivated chips were fitted with a two-component TLS loss model at a fixed temperature, hence reflecting solely the power dependence of the TLSs loss behavior [14–17]. The typical result of a fitting of eq. 4 to selected resonator data is shown in Fig. S9, for a passivated and an un-passivated resonator; all the fitting parameters ($b_i$, $n_i$, $\tilde{\delta}_{i,TLS}$) were then derived from the fitting routine. Indices $i= 1,2$ and $i= 1,3$ were used for un-passivated and passivated resonators, respectively, signifying one common loss channel and a second, distinct channel for either resonator. The fitting routines were performed within a 95% confidence interval for all the resonators. Further, Kendall correlation coefficients were calculated using OriginLab software for all fitting parameters; the corresponding correlation matrix (Fig. S10) shows no strong correlation (<0.7) between any two variables, suggesting that there was no redundant fitting parameter in the procedure.

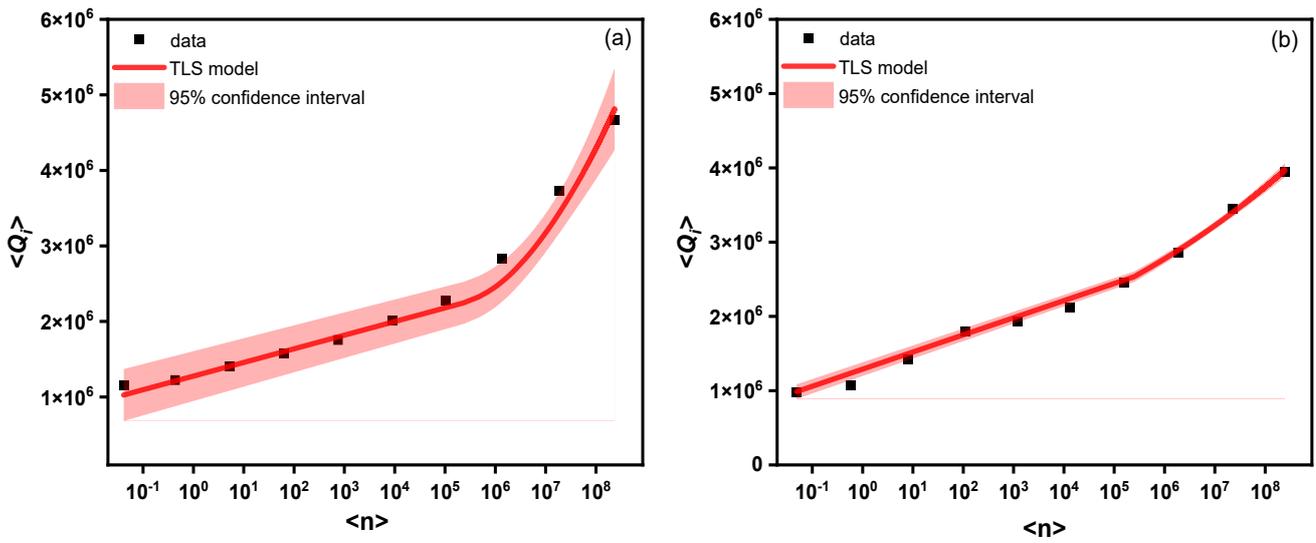

***Figure S9*** *Example fitting results of the two-component TLS model to $Q_i$ vs. $<n>$ data, for resonators at a resonance frequency of 6.1GHz: (a) a passivated resonator at day 0, (b) a passivated resonator at day 6.*

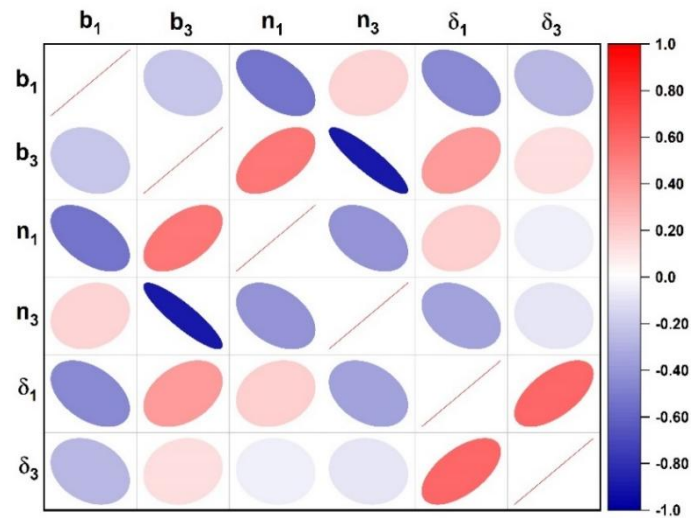

***Figure S10*** *An exemplary correlation matrix between different fitting parameters in the two-component TLS model used to fit the Q-n behavior of resonators (passivated resonator on day 0). A positive slope of the major axis of the ellipses indicates a positive correlation and vice versa; the ratio of the major axis with respect to the minor axis for of the ellipses represents the extent of correlation (smaller minor axis --> stronger correlation). The color code equally shows the magnitude of correlation, ranging from red (strong) to blue (weak). No strong correlation (<0.7) is observed for the different fitting parameters.*

**Additional resonator data**

In addition to the data presented in the main manuscript, two more sets of resonator samples were measured and their data fitted with the two-TLS model. Figure S11 shows selected $<Q_i>$ vs. $<n>$ data for these sets, with characteristic photon numbers ($n_i$) extracted by fitting. Set-2 shows an almost similar $Q_i$ in the low photon regime for the fresh (day0) passivated and un-

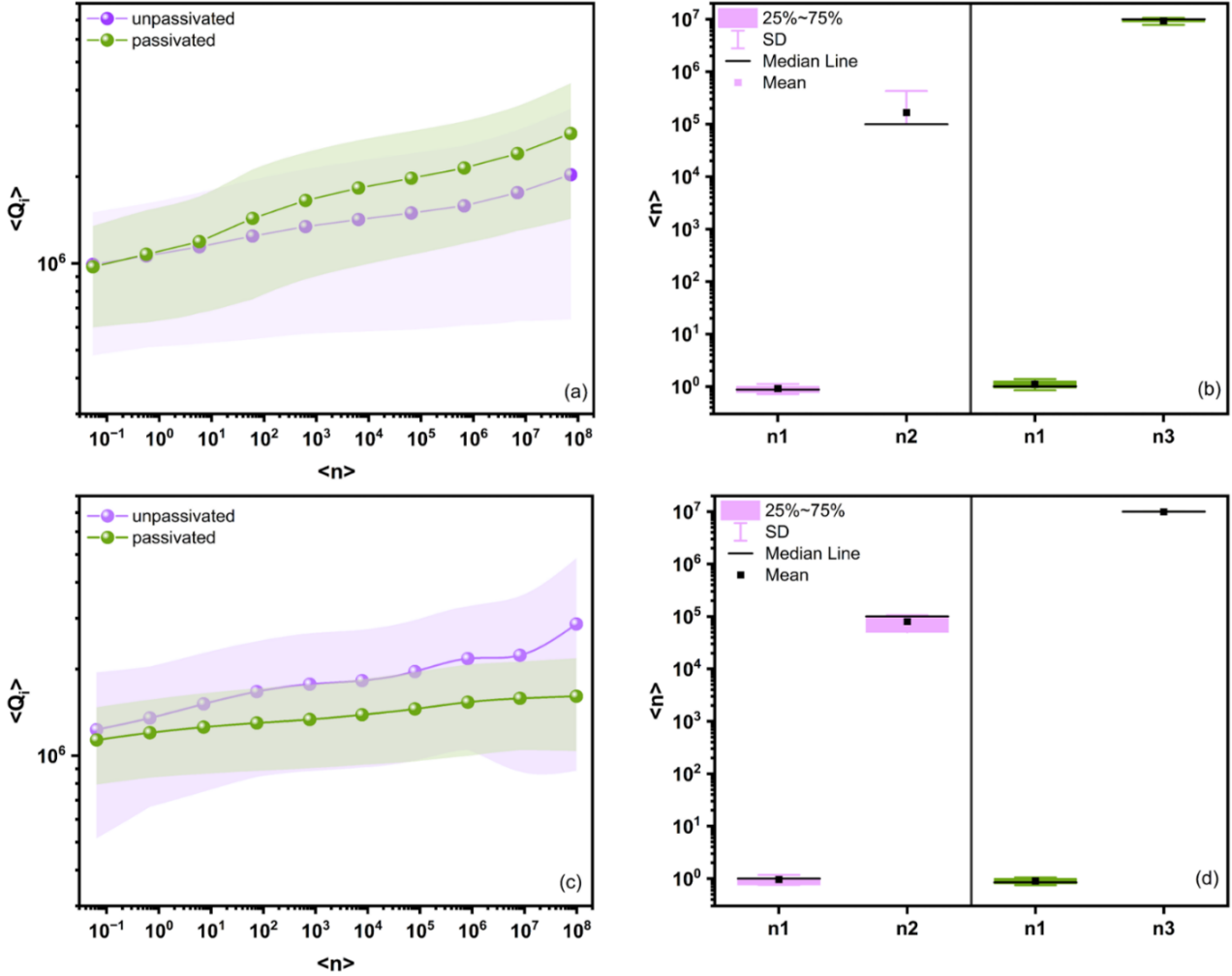

*Figure S11 (a), (c) Mean internal quality factor ($Q_i$) for Nb resonators against mean photon number for fresh (day0) set2 and set3 respectively, the shaded region in the plot denotes the standard deviation of the data. (b), (d) Characteristic critical photon numbers extracted from the TLS model: $n_1$, $n_2$ for un-passivated resonators (purple) and $n_1$, $n_3$ for passivated resonators (green) for set2 and set3, respectively.*

passivated resonators, whereas the passivated resonators show higher $Q_i$ values in the high photon number regime. For set-3 in the low photon number regime, the trend is similar however, compared to set-2 an opposite trend in the high photon number regime is observed. Similar to set-1 (main manuscript) sets 2 and 3 both clearly indicate the presence of two discrete loss channels in the un-passivated and passivated resonators, corresponding to two distinct critical photon numbers ($n_1$ ~$10^0$, $n_2$ ~$10^5$ for un-passivated, and $n_1$ ~$10^0$ and $n_3$ ~$10^7$ for passivated resonators). With $n_3$ being notably apart ($n_3>n_2$) from the loss channel observed in un-passivated resonators ($i$=2), the loss component with the higher critical photon number $n_2$ is assigned to the regrowth of $Nb_2O_5$ ($\tilde{\delta}_{2,TLS}= \tilde{\delta}_{Nb2O5}$) and the loss channel corresponding to a high critical photon number ($n_3$~$10^7$) is ascribed to the passivating SAM on the Nb resonators ($\tilde{\delta}_{3,TLS}= \tilde{\delta}_{SAM}$).

The extracted loss for SAMs ($\tilde{\delta}_{SAM}$) for sets 2, 3 is also found to be in the range of 5-7x10$^{-7}$, similar to set 1. It should be noted though that these sets of samples show a slightly different $Q$-$n$ behavior compared to set 1; it is attributed to an overall lower $Q_i$ for this set of resonators even for freshly BOE etched resonators, which we assign to a different (less optimized) fabrication process in that case. However, the dominant loss channels are similar to set-1, resulting in similar critical photon numbers.

## References

1. Zhao, D.-D. *et al.* Optical constants of e-beam evaporated and annealed Nb 2 O 5 thin films with varying thickness. *J. Phys. D: Appl. Phys.* **49,** 265304; 10.1088/0022-3727/49/26/265304 (2016).

2. Atul N. Parikh, David L. Allara, Issam Ben Azouz & Francis Rondelez. An Intrinsic Relationship between Molecular Structure in Self-Assembled n-Alkylsiloxane Monolayers and Deposition Temperature. *J. Phys. Chem.* **98,** 7577–7590 (1994).

3. Altoé, M. V. P. *et al.* Localization and Mitigation of Loss in Niobium Superconducting Circuits. *PRX Quantum* **3,** 20312; 10.1103/PRXQuantum.3.020312 (2022).

4. Morgan, D. J. XPS insights: Asymmetric peak shapes in XPS. *Surface & Interface Analysis* **55,** 567–571; 10.1002/sia.7215 (2023).

5. Darlinski, A. & Halbritter, J. Angle‐resolved XPS studies of oxides at NbN, NbC, and Nb surfaces. *Surface & Interface Analysis* **10,** 223–237; 10.1002/sia.740100502 (1987).

6. Demchenko, I. N. *et al.* Effect of argon sputtering on XPS depth-profiling results of Si/Nb/Si. *Journal of Electron Spectroscopy and Related Phenomena* **224,** 17–22; 10.1016/j.elspec.2017.09.009 (2018).

7. Cumpson, P. J. & Seah, M. P. Elastic Scattering Corrections in AES and XPS. II. Estimating Attenuation Lengths and Conditions Required for their Valid Use in Overlayer/Substrate Experiments. *Surf. Interface Anal.* **25,** 430–446; 10.1002/(SICI)1096-9918(199706)25:6<430::AID-SIA254>3.0.CO;2-7 (1997).

8. Grundner, M. & Halbritter, J. XPS and AES studies on oxide growth and oxide coatings on niobium. *Journal of Applied Physics* **51,** 397–405; 10.1063/1.327386 (1980).

9. Khalil, M. S., Stoutimore, M. J. A., Wellstood, F. C. & Osborn, K. D. An analysis method for asymmetric resonator transmission applied to superconducting devices. *Journal of Applied Physics* **111,** 54510; 10.1063/1.3692073 (2012).

10. Probst, S., Song, F. B., Bushev, P. A., Ustinov, A. V. & Weides, M. Efficient and robust analysis of complex scattering data under noise in microwave resonators. *The Review of scientific instruments* **86,** 24706; 10.1063/1.4907935 (2015).

11. Tai, T., Cai, J. & Anlage, S. M. Anomalous Loss Reduction Below Two‐Level System Saturation in Aluminum Superconducting Resonators. *Adv Quantum Tech* **7,** 2200145; 10.1002/qute.202200145 (2024).

12. Burnett, J., Bengtsson, A., Niepce, D. & Bylander, J. Noise and loss of superconducting aluminium resonators at single photon energies. *J. Phys.: Conf. Ser.* **969,** 12131; 10.1088/1742-6596/969/1/012131 (2018).

13. McRae, C. R. H. *et al.* Materials loss measurements using superconducting microwave resonators. *The Review of scientific instruments* **91,** 91101; 10.1063/5.0017378 (2020).

14. Kirsh, N., Svetitsky, E., Burin, A. L., Schechter, M. & Katz, N. Revealing the nonlinear response of a tunneling two-level system ensemble using coupled modes. *Phys. Rev. Materials* **1,** 12601; 10.1103/PhysRevMaterials.1.012601 (2017).

15. Verjauw, J. *et al.* Investigation of Microwave Loss Induced by Oxide Regrowth in High-Q Niobium Resonators. *Phys. Rev. Applied* **16,** 14018; 10.1103/PhysRevApplied.16.014018 (2021).

16. Faoro, L. & Ioffe, L. B. Internal loss of superconducting resonators induced by interacting two-level systems. *Physical review letters* **109,** 157005; 10.1103/PhysRevLett.109.157005 (2012).

17. Schechter, M. & Stamp, P. C. E. Inversion symmetric two-level systems and the low-temperature universality in disordered solids. *Phys. Rev. B* **88,** 174202; 10.1103/PhysRevB.88.174202 (2013).